\documentclass[]{aa}  

\usepackage{graphicx,subcaption}
\usepackage{xcolor}
\usepackage{tikz}
\usepackage{xspace}
\usepackage[hidelinks]{hyperref}
\usepackage{txfonts}

\usepackage{float,subfloat}

\newcommand{\micron}{\rm \mu m}
\newcommand{\ergs}{\rm erg\,s^{-1}}
\newcommand{\Oiii}{[O\,{\sc iii}]\xspace}
\newcommand{\Hei}{He\,{\sc i}\xspace}
\newcommand{\Oi}{O\,{\sc i}\xspace}
\newcommand{\Nii}{[N\,{\sc ii}]\xspace}
\newcommand{\Hb}{{\rm H}$\beta$\xspace}

\newcommand{\Msun}{\rm M_\odot}

\newcommand{\kms}{\,\rm km\,s^{-1}}
\newcommand{\re}{r_{\rm e,maj}}
\newcommand{\rec}{r_{\rm e}}

\newcommand{\Mbh}{M_{\rm BH}}

\newcommand{\Ha}{H$\alpha$\xspace}
\newcommand{\Pad}{Pa-$\delta$\xspace}
\newcommand{\Pag}{Pa-$\gamma$\xspace}

\newcommand{\ruby}{\textit{The Cliff}\xspace} 
\newcommand{\zspec}{z_{\rm spec}}
\newcommand{\zphot}{z_{\rm phot}}
\newcommand{\pix}{\rm pix}
\newcommand{\prospector}{\texttt{Prospector}\xspace}

\newcommand{\Av}{A_{\rm V}}
\newcommand{\cloudy}{\texttt{Cloudy}\xspace}
\newcommand{\mombh}{MoM-BH*-1\xspace}

\begin{document}

% Anna de Graaff, Hans-Walter Rix, Rohan P. Naidu, Ivo Labbe, Bingjie Wang, Joel Leja, Jorryt Matthee, Harley Katz, Jenny E. Greene, Raphael E. Hviding, Josephine Baggen, Rachel Bezanson, Leindert A. Boogaard, Gabriel Brammer, Pratika Dayal, Pieter van Dokkum, Andy D. Goulding, Michaela Hirschmann, Michael V. Maseda, Ian McConachie, Tim B. Miller, Erica Nelson, Pascal A. Oesch, David J. Setton, Irene Shivaei, Andrea Weibel, Katherine E. Whitaker, Christina C. Williams

\title{A remarkable Ruby: Absorption in dense gas, rather than evolved stars, drives the extreme Balmer break of a Little Red Dot at $z=3.5$}

   \author{Anna de Graaff \inst{\ref{i1}}\thanks{degraaff@mpia.de}
         \and Hans-Walter Rix \inst{\ref{i1}}
         \and Rohan P. Naidu \inst{\ref{i_mit}}\thanks{NASA Hubble Fellow}
         \and Ivo Labb\'e \inst{\ref{i_swin}}
         \and Bingjie Wang \inst{\ref{i4}, \ref{i5}, \ref{i6}}
         \and Joel Leja \inst{\ref{i4}, \ref{i5}, \ref{i6}}
         \and Jorryt Matthee \inst{\ref{i_ista}}
         \and Harley Katz \inst{\ref{i_chicago}}
         \and Jenny E. Greene \inst{\ref{i_princeton}}
         \and Raphael E. Hviding \inst{\ref{i1}}
         \and Josephine Baggen \inst{\ref{i_yale}}
         \and Rachel Bezanson \inst{\ref{i_pitt}}
         \and Leindert A. Boogaard\inst{\ref{i_strw}}
         \and Gabriel Brammer\inst{\ref{i2},\ref{i_NBI}}        
         \and Pratika Dayal \inst{\ref{kap}}
         \and Pieter van Dokkum \inst{\ref{i_yale}}         
         \and Andy D. Goulding \inst{\ref{i_princeton}}
         \and Michaela Hirschmann \inst{\ref{epfl}}
         \and Michael V. Maseda\inst{\ref{i3}}
         \and Ian McConachie\inst{\ref{i3}}
         \and Tim B. Miller \inst{\ref{i_ciera}}
         \and Erica Nelson \inst{\ref{cub}}
         \and Pascal A. Oesch \inst{\ref{i_ge},\ref{i2},\ref{i_NBI}}
         \and David J. Setton \inst{\ref{i_princeton}}\thanks{Brinson Prize Fellow}
        \and Irene Shivaei \inst{\ref{cab}}
         \and Andrea Weibel \inst{\ref{i_ge}}
         \and Katherine E. Whitaker\inst{\ref{i_umass},\ref{i2}}
        \and Christina C. Williams \inst{\ref{NOIRLab}}
 }

   \institute{Max-Planck-Institut f\"ur Astronomie, K\"onigstuhl 17, D-69117 Heidelberg, Germany\label{i1}
    \and 
    MIT Kavli Institute for Astrophysics and Space Research, 77 Massachusetts Ave., Cambridge, MA 02139, USA\label{i_mit}
    \and
    Centre for Astrophysics and Supercomputing, Swinburne University of Technology, Melbourne, VIC 3122, Australia\label{i_swin}   \and
    Department of Astronomy \& Astrophysics, The Pennsylvania State University, University Park, PA 16802, USA\label{i4}
    \and 
    Institute for Computational \& Data Sciences, The Pennsylvania State University, University Park, PA 16802, USA\label{i5}
    \and 
    Institute for Gravitation and the Cosmos, The Pennsylvania State University, University Park, PA 16802, USA\label{i6}
    \and 
    Institute of Science and Technology Austria (ISTA), Am Campus 1, 3400 Klosterneuburg, Austria\label{i_ista}  
    \and 
    Department of Astronomy \& Astrophysics, University of Chicago, 5640 S Ellis Avenue, Chicago, IL 60637, USA\label{i_chicago}    
     \and 
    Department of Astrophysical Sciences, Princeton University, 4 Ivy Lane, Princeton, NJ 08544, USA\label{i_princeton}
    \and
    Astronomy Department, Yale University, 219 Prospect St, New Haven, CT 06511, USA\label{i_yale} 
    \and 
    Department of Physics and Astronomy and PITT PACC, University of Pittsburgh, Pittsburgh, PA 15260, USA\label{i_pitt}
     \and 
     Leiden Observatory, Leiden University, PO Box 9513, NL-2300 RA Leiden, The Netherlands\label{i_strw}
    \and 
    Cosmic Dawn Center (DAWN), Copenhagen, Denmark\label{i2} 
    \and 
    Niels Bohr Institute, University of Copenhagen, Jagtvej 128, Copenhagen, Denmark\label{i_NBI} 
     \and 
     Kapteyn Astronomical Institute, University of Groningen, PO Box 800, 9700 AV Groningen, The Netherlands\label{kap}
    \and 
    Institute of Physics, Laboratory for Galaxy Evolution, Ecole Polytechnique Federale de Lausanne, Observatoire de Sauverny, Chemin Pegasi 51, 1290 Versoix, Switzerland\label{epfl}
    \and 
    Department of Astronomy, University of Wisconsin-Madison, 475 N. Charter St., Madison, WI 53706 USA\label{i3}  
    \and 
    Center for Interdisciplinary Exploration and Research in Astrophysics (CIERA), Northwestern University,1800 Sherman Ave, Evanston, IL 60201, USA\label{i_ciera}
    \and
    Department for Astrophysical and Planetary Science, University of Colorado, Boulder, CO 80309, USA \label{cub}
    \and 
    Department of Astronomy, University of Geneva, Chemin Pegasi 51, 1290 Versoix, Switzerland\label{i_ge} 
     \and 
     Centro de Astrobiolog{\'i}a (CAB), CSIC-INTA, Ctra. de Ajalvir km 4, Torrejón de Ardoz, E-28850, Madrid, Spain\label{cab}    
    \and 
     Department of Astronomy, University of Massachusetts, Amherst, MA 01003, USA\label{i_umass}
     \and
     NSF’s National Optical-Infrared Astronomy Research Laboratory, 950 North Cherry Avenue, Tucson, AZ 85719, USA\label{NOIRLab}
    }

\abstract{
\noindent 

The origin of the rest-optical emission of compact, red, high-redshift sources known as `little red dots' (LRDs) poses a major puzzle. If interpreted as starlight, it would imply that LRDs would constitute the densest stellar systems in the Universe. However, alternative models suggest active galactic nuclei (AGN) may instead power the rest-optical continuum. 
Here, we present JWST/NIRSpec, NIRCam and MIRI observations from the RUBIES and PRIMER programs of \ruby: a bright LRD at $z=3.55$ with an exceptional Balmer break, twice as strong as that of any high-redshift source previously observed. The spectra also reveal broad Hydrogen (\Ha $\rm FWHM\sim1500\,\kms$) and \Hei emission, but no significant metal lines. We demonstrate that massive evolved stellar populations cannot explain the observed spectrum, even when considering unusually steep and strong dust attenuation, or reasonable variations in the initial mass function. 
Moreover, the formally best-fit stellar mass and compact size ($M_*\sim10^{10.5}\,\Msun,\ r_{\rm e}\sim40\,$pc) would imply densities at which near-monthly stellar collisions might lead to significant X-ray emission. 
We argue that the Balmer break, emission lines, and \Ha absorption line are instead most plausibly explained by a `\emph{black hole star}' (BH*) scenario, in which dense gas surrounds a powerful ionising source. In contrast to recently proposed BH* models of dust-reddened AGN, we show that spectral fits in the rest UV to near-infrared favour an intrinsically redder continuum over strong dust reddening. This may point to a super-Eddington accreting massive black hole or, possibly, the presence of (super)massive stars in a nuclear star cluster. \ruby is the clearest evidence to date that at least some LRDs are not ultra-dense, massive galaxies, and are instead powered by a central ionising source embedded in dense, absorbing gas.
}

   \keywords{Galaxies: evolution -- Galaxies: active -- Galaxies: kinematics and dynamics -- Galaxies: stellar content}

\titlerunning{A Balmer Cliff}
\authorrunning{de Graaff et al.}

\maketitle

\clearpage
\section{Introduction} \label{sec:intro}

Among the most debated discoveries with the James Webb Space Telescope (JWST) are the high-redshift compact red sources referred to as `little red dots' (LRDs). Although originally coined as a term for sources that were solely selected on their broad symmetric \Ha lines, they turned out to also be ubiquitously red and compact \citep{Matthee2024}, leading to a variety of definitions that are currently in use in the literature. Typically, photometric selections of LRDs require that sources are (1) compact in the rest-optical and (2) have `v-shaped' rest UV-to-optical spectral energy distributions (SEDs), or simply extremely red rest-frame optical colours \citep[e.g.][]{Labbe2023b,Kocevski2024,Kokorev2024,Akins2024}. Follow-up spectroscopy has revealed luminous and broad Balmer emission lines in a large fraction of these systems, as well as blue rest-UV and red rest-optical continua \citep{Greene2024,Kocevski2024,Setton2024,Wang2024b,Hviding2025}. The primary subject of debate now is whether these broad emission lines and v-shaped continua are powered by accreting massive black holes, likely constituting a new class of active galactic nuclei (AGN), or whether the SED properties can instead be attributed to (extremely) massive stellar populations \citep[e.g.][]{Labbe2023,Greene2024,Wang2024b,Baggen2024,Guia2024,Merida2025}.

Determining whether the SEDs of LRDs are dominated by stars or AGN is a major challenge, even with high-quality rest-optical spectroscopy and photometry ranging from X-ray to radio wavelengths. Although broad ($\rm FWHM\gtrsim1000\,\kms$) symmetric Balmer emission lines -- in combination with narrow forbidden lines -- are normally a key signature of broad-line AGN, LRDs lack other typical characteristics of AGN. Notably, LRDs show extremely weak, if any, X-ray emission \citep[e.g.][]{Ananna2024,Maiolino2023,Yue2024}. Observations with JWST/MIRI have also revealed surprisingly blue colours in the rest-frame near- and mid-infrared \citep{Wang2024a,Williams2024,Perez2024,Setton2025}, which implies a lack of hot dust emission from AGN tori. Moreover, while some LRDs do show AGN signatures at rest-UV wavelengths in the form of strong emission lines \citep{Labbe2024,Akins2024b}, the UV emission of many others can be well modelled by a (low-mass) star-forming galaxy \citep[e.g.][]{Maiolino2023,Killi2023,Barro2024}.

The origin of the rest-optical continuum emission is perhaps most contentious. If dominated by the emission from a (strongly) dust-attenuated AGN, the implied host galaxy mass is small, resulting in a black hole to host mass ratio that lies far above well-established scaling relations in the local Universe \citep[e.g.][]{Maiolino2023,Pacucci2023,Furtak2024}, although this may in part be driven by uncertainties in the black hole and galaxy masses as well as selection effects \citep[e.g.][]{Li2024,Lupi2024,Wang2024b}. If confirmed, such a large population of overmassive black holes may have important implications for the seed masses and growth mechanisms of early black holes \citep[e.g.][]{Kokorev2023,Bogdan2024,Goulding2023}, and would be in tension with current cosmological hydrodynamical simulations \citep[e.g.][]{Habouzit2025,Matthee2024b}. Intriguingly, however, the transition in the v-shaped SED from blue to red has been shown to occur at the Balmer limit for $\sim50\%$ of LRDs \citep{Setton2024}, and some LRDs show Balmer breaks that are as strong as those observed in high-redshift post-starburst galaxies \citep[e.g.][]{Labbe2024,Wang2024b,Ma2024}. At face value, it therefore appears reasonable to interpret the rest-optical as being dominated by emission from an evolved stellar population in a host galaxy instead of by an AGN. 

However, such a stellar mass-dominated interpretation leads to great tension with galaxy formation models. The stellar masses of LRDs obtained from SED fitting are typically high \citep[$M_*\sim10^{10}\,\Msun$ under the assumption of a Milky Way-like initial mass function; e.g. ][]{Akins2024}, with some galaxies being as massive as $M_*\sim10^{11}\,\Msun$ by $z\sim 7-8$ \citep{Labbe2023,Wang2024b}. Coupled with the relatively high number densities of LRDs (e.g. \citealt{Kokorev2024}; and thus modest implied halo masses, e.g. \citealt{Pizzati2024}), this would imply very high conversion rates from available baryons in haloes into stellar mass. The most luminous LRDs are estimated to be so massive that the extreme galaxy formation efficiencies required for their formation are difficult to accommodate within the $\Lambda$CDM model \citep{MBK2023}. Moreover, because of the very compact morphologies of LRDs, the high masses also imply exceptionally high stellar mass densities, which exceed the maximum expected stellar densities from observations of dense star clusters and star formation models \citep{Baggen2023,Baggen2024,Guia2024,Labbe2024,Ma2024}.

To bridge the predominantly AGN or predominantly stellar scenarios, SED modelling efforts have therefore focused on multi-component fitting of photometric \citep[e.g.][]{Barro2024,Perez2024,Juodzbalis2024,Leung2024} and spectrophotometric data \citep[][]{Harikane2023,Labbe2024,Ma2024,Maiolino2023,Wang2024a,Wang2024b}. Generally, including some dust-reddened AGN component lowers the inferred stellar mass and stellar mass density (by up to $\sim 2$\,dex), reducing some of the aforementioned tension with galaxy formation models. However, several of these SED models, especially when fitting spectra, require an uncomfortable level of tuning: e.g., strong AGN emission must consistently arise \emph{just} redward of the Balmer break, and an unusually steep dust extinction law is needed to match the spectral shape \citep{Wang2024b,Ma2024}.  

Most recently, the interpretation that the observed Balmer break must originate from stellar emission has been called into question \citep{Inayoshi2025,Ji2025}. Instead, these studies suggest that extremely dense gas in the close vicinity of an AGN can resemble the conditions of ($\sim10^4$\,K) stellar atmospheres, and hence result in spectra that mimic the SEDs of high-redshift quiescent galaxies (although the AGN models of \citealt{Ji2025} require substantially greater dust attenuation). This is further motivated by the discovery of strong absorption features in the Balmer and \Hei emission lines, which point to the presence of absorbing dense gas clouds \citep{Matthee2024,Wang2024a,Juodzbalis2024,DEugenio2025}. Moreover, such an AGN-dominated scenario would also be in line with recent clustering measurements which independently suggest low host galaxy masses in a sample of fainter LRDs \citep[][]{Matthee2024b}.

Critically, however, it has so far been difficult to robustly differentiate between these various models, especially for the brightest LRDs with strong Balmer breaks. That is, models in which a massive galaxy dominates the SED often fit the rest-frame UV to near-IR spectrophotometric data equally well as multi-component galaxy+AGN models or the newly proposed non-stellar Balmer break model. Arguments against or in favour of specific models have therefore relied on other evidence, such as the X-ray non-detections, emission line kinematics, morphology, or the degree of model fine-tuning required.

In particular, conclusively ruling out the presence of a massive, ultra-dense stellar population dominating the rest-optical has been surprisingly difficult. Although this scenario requires invoking high star formation efficiencies, its simplicity is appealing: the SED, including the lack of X-ray emission and hot dust in the mid-IR, can largely be explained with standard stellar population models, and the broad emission line kinematics have been suggested to arise naturally from the very high stellar mass density \citep{Baggen2024}.

In this paper we present a luminous LRD at $\zspec=3.548$ from the Red Unknowns: Bright Infrared Extragalactic Survey \citep[RUBIES;][]{deGraaff2024d}, named \ruby, as JWST/NIRSpec spectroscopy reveals an exceptionally strong Balmer break. We rigorously assess the possibility that this remarkable source is stellar in origin, by testing a range of stellar and AGN models and by exploring its implied stellar dynamical properties. We will demonstrate that we can, for the first time, robustly exclude a high stellar mass and high stellar density origin for a LRD with a strong Balmer break. The photometric and spectroscopic data are described in Section~\ref{sec:data}, and measurements of the emission line and morphological properties are presented in Sections~\ref{sec:spec} and \ref{sec:morph}, respectively. We present our stellar population modelling in Section~\ref{sec:sed_models}, and explore the consequences of these models on the dust properties and galaxy dynamical properties in Sections~\ref{sec:dust} and \ref{sec:collisions}. We also assess possible variations in the stellar initial mass function in Section~\ref{sec:imf}. Finally, we propose in Section~\ref{sec:cloudy} that the preferred model for this source is likely that of an accreting massive black hole embedded in high-density gas, and then discuss the intrinsic spectrum of the ionising source. Our observations and findings are summarised in Section~\ref{sec:conclusions}. Where relevant, we adopt the best-fit cosmological parameters from the WMAP 9 yr results: $H_{0}=69.32$ ${\rm km \,s^{-1} \,Mpc^{-1}}$, $\Omega_{\rm m}=0.2865$, and $\Omega_{\Lambda}=0.7135$ \citep{Hinshaw2013}.

\section{Data}\label{sec:data}

\subsection{Imaging}\label{sec:imaging}

The primary object of this paper lies in the UDS field (R.A. 02:17:38.58; Dec. -05:07:46.79), and was selected (see Section~\ref{sec:data-spectroscopy}) from public JWST/NIRCam and MIRI imaging obtained as part of the PRIMER Survey (GO-1837, PI: Dunlop). The PRIMER imaging covers 8 NIRCam filters (F090W, F115W, F150W, F200W, F277W, F356W, F410M, F444W; see e.g. \citealt{Donnan2024} for details) as well as 2 MIRI filters (F770W, F1800W).

We use the publicly available image mosaics from the Dawn JWST Archive (DJA, version 7.2), which were reduced with the \texttt{grizli} software \citep{grizli} as described in \citet{Valentino2023}. These mosaics have a pixel scale of $0.04\arcsec$; for the purposes of morphological analysis in Section~\ref{sec:morph}, we also use a custom image mosaic for the F200W filter with a better pixel sampling of $0.02\arcsec$ (as was also used in \citealt{Weibel2024b}).

\begin{figure*}
    
    \centering
    \includegraphics[width=\linewidth]{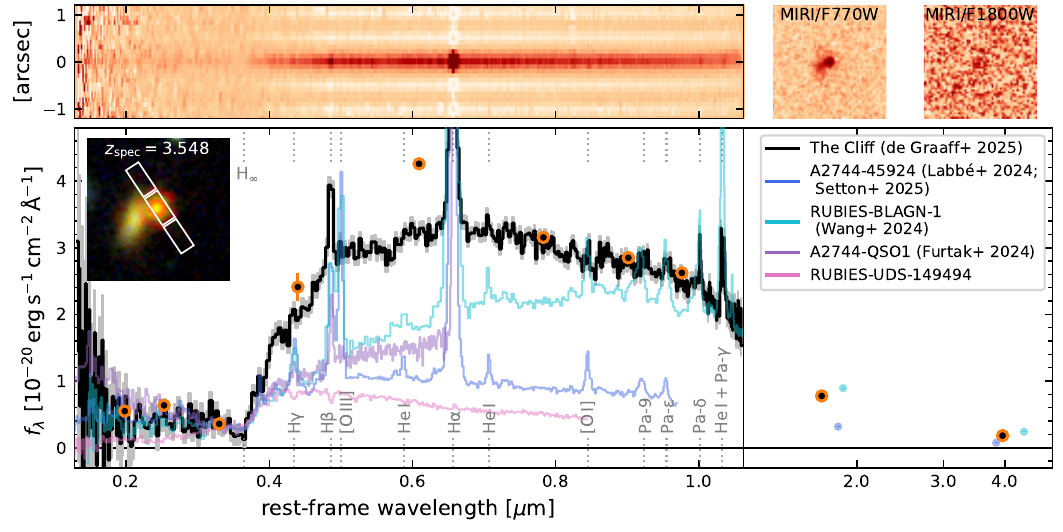}
    \caption{NIRSpec/PRISM spectrum of \ruby (RUBIES-UDS-154183) at $\zspec=3.548$. Orange points show NIRCam and MIRI photometry spanning an observed wavelength range of $0.9-18\,\micron$; the inset colour image was constructed using the NIRCam F115W, F277W and F444W filters and shows the location of the NIRSpec microshutters. Coloured lines are the NIRSpec/PRISM spectra (rescaled using the median flux at rest-frame $[3200-3700]\,\AA$) of four sources with strong Balmer breaks: two of the most luminous LRDs in the literature \citep[$\zspec\sim3-5$][]{Labbe2024,Wang2024a} and their MIRI detections \citep{Setton2025}, the triple-imaged LRD at $\zspec=7.04$ of \citet{Furtak2024}, and a massive post-starburst galaxy at $\zspec=4.62$ (a medium-resolution NIRSpec spectrum of which was presented by \citealt{Carnall2024}). Dotted lines indicate the locations of strong emission line features in the different spectra as well as the Balmer limit. \ruby shows an exceptionally strong Balmer break, a declining SED in the rest near-infrared ($\sim1-4\,\micron$), strong H and \Hei emission lines, but (in comparison to the two luminous LRDs) very weak metal lines.}
    \label{fig:spec}
\end{figure*}

The inset of Figure~\ref{fig:spec} shows a colour image from the F115W, F277W and F444W filters, revealing a compact red source (our target) as well as a blue neighbouring source. We obtain a photometric redshift $\zphot=2.78^{+0.17}_{-0.05}$ of this neighbour from the catalogue of \citet{Weibel2024}, substantially lower than the redshift of \ruby. {Although it is not physically associated, we do find that the circular aperture photometry at short wavelengths is contaminated by the foreground source, because the foreground source has a very blue colour and the photometry was measured from mosaics that were matched to the coarser spatial resolution of the F444W filter.}

We therefore perform simultaneous S\'ersic profile fitting with \texttt{pysersic} \citep{Pasha2023} in order to obtain deblended fluxes. Each source is modelled as a single S\'ersic profile, and convolved with the empirical {point spread functions (PSFs)} of \citet{Weibel2024}. We fit the available NIRCam filters using uniform priors for all parameters, restricting the S\'ersic index to the range $[0.65,4]$ and the effective radius (major axis) to $[0.5,100]\,\pix$, and using the No U-turn sampler of \texttt{numpyro} to sample the posterior distributions \citep{hoffman2014,phan2019}. {The NIRCam images, maximum a posteriori (MAP) models, and residual images can be found in Appendix~\ref{sec:apdx_photometry}. As described in Appendix~\ref{sec:apdx_photometry}, we find that the S\'ersic profile model of \ruby strongly dominates the flux at the position of the NIRSpec microshutter. In other words, contamination from the nearby neighbour does not significantly contribute to the observed NIRSpec spectrum or affect any of the conclusions in this work.}

The MIRI images are shallower and we therefore only fit for the flux, informed by the fit in the NIRCam/F200W filter and using the empirical MIRI PSF models of \citet{Libralato2024}. We model the foreground source as a single S\'ersic profile, and set truncated (at $\pm 3\sigma$) Gaussian priors for the model parameters based on the posteriors of the F200W fit. Because the posterior of the effective radius of \ruby is at the border of the prior ($\re\approx 0.5\,\pix$), we fit a simpler point source model instead of a S\'ersic model for the MIRI filters (with a prior on the position from the F200W posteriors). The pixel scale of the MIRI image mosaics ($0.04\arcsec$) is substantially smaller than the pixel scale of the MIRI instrument ($0.11\arcsec$), and the pixels are therefore highly correlated. This correlated noise is not accounted for by \texttt{pysersic}, which only uses the variance image. To estimate uncertainties on the derived fluxes, we therefore perform point source photometry with \texttt{pysersic} in random empty areas in the vicinity of \ruby, masking bright sources. We take the standard deviation of these random measurements as the uncertainty on the point source flux, which is approximately twice higher than the formal uncertainties of \texttt{pysersic}. A table with all NIRCam and MIRI fluxes{, as well as the images, morphological models and residual images, are} provided in Appendix~\ref{sec:apdx_photometry}.

\subsection{Spectroscopy}\label{sec:data-spectroscopy}

Observations with the JWST/NIRSpec microshutter array (MSA) were obtained in July 2024 as part of RUBIES (GO-4233; PIs: de Graaff \& Brammer). \ruby (RUBIES-UDS-154183) was selected with high priority for its very red colour ($\rm F150W-F444W\approx3.9$) and promoted to the highest `Priority Class 0' because of its extremely compact morphology and strong photometric Balmer break. 

The details of the RUBIES priority classes and observing strategy can be found in \citet{deGraaff2024d}. Briefly, we observe with both the low-resolution PRISM/Clear ($0.6-5.3\,\micron$) and medium-resolution G395M/F290LP ($2.9-5.2\,\micron$) disperser/filter combinations, using a 3-point nodding pattern and total exposure time of 48\,min per disperser. The data were reduced with the \texttt{msaexp} software and correspond to version 3 of the DJA, as described in \citet{Heintz2024} and \citet{deGraaff2024d}. Notably, this version provides two types of background subtraction for the PRISM spectra: a local subtraction from the image differences between the nodded exposures, and a global background subtraction obtained from empty sky shutters in the mask. For consistency, because only the local background subtraction is available for the G395M spectra, we use the local background subtraction for both dispersers throughout (although we note that in practice this makes no significant difference to any of the results in this paper).

\subsection{X-ray}\label{sec:xray}

We obtain public Chandra X-ray data from the X-UDS survey \citep{Kocevski2018}. The location of \ruby was observed with effective exposure times of 573.9\,ks ($<2\,$keV) and 609.2\,ks ($2-7\,$keV). Following \citet{Wang2024a}, we obtain $3\sigma$ upper limits on the X-ray luminosities: $L_X <1.9\times10^{44}\,{\rm erg\,s^{-1}}$ (rest $1-4\,$keV), $L_X <3.5\times10^{42}\,{\rm erg\,s^{-1}}$ (rest $4.5-9\,$keV), $L_X <6.7\times10^{43}\,{\rm erg\,s^{-1}}$ (rest $10-30\,$keV).

\section{Spectral properties}\label{sec:spec}

\begin{figure}
    \centering
    \includegraphics[width=\linewidth]{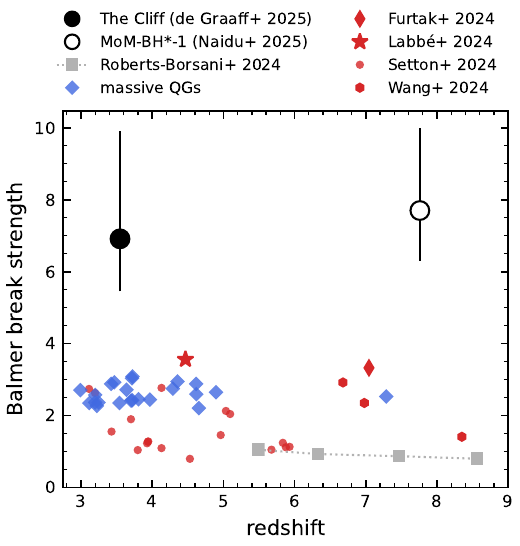}
    \caption{Balmer break strength, measured as the ratio of the mean flux density in the rest-frame wavelength ranges $[3620,3720]\,\AA$ and $[4000,4100]\,\AA$ from public JWST/NIRSpec data. Grey squares show the median values of stacks of star-forming galaxies from \citet{RobertsBorsani2024}; red markers represent a compilation of LRDs \citep{Furtak2024,Labbe2024,Setton2024,Wang2024b}; blue markers show a large sample of massive quiescent galaxies \citep[compiled from][]{Barrufet2025,Carnall2024,deGraaff2024c,Glazebrook2024,Nanayakkara2024,Weibel2024b}. The recently discovered LRD of \citet{Naidu2025}, \mombh, is a higher-redshift analogue of \ruby and discussed in Section~\ref{sec:cloudy}.}
    \label{fig:BB_zspec}
\end{figure}

The 2D PRISM spectrum of \ruby and its 1D extraction are shown in Figure~\ref{fig:spec}. The NIRCam and MIRI fluxes are also overplotted, and demonstrate that the absolute flux calibration of the NIRSpec spectrum is in good agreement with the photometry. We further show the rescaled NIRSpec/PRISM spectra of four sources at $\zspec>3$ with very strong Balmer breaks: A2744-45924 \citep[$\zspec=4.47$][]{Labbe2024} and RUBIES-BLAGN-1 \citep[$\zspec=3.10$][]{Wang2024a} are two of the most luminous LRDs found to date; the triply-imaged LRD A2744-QSO1 has the strongest Balmer break observed so far at $z>6$ \citep[$\zspec=7.04$][]{Furtak2024}; RUBIES-UDS-149494 is a massive quiescent galaxy at $\zspec=4.62$ (a medium-resolution spectrum of which was also presented in \citealt{Carnall2024} with ID PRIMER-EXCELS-117560). Comparing to the other LRDs, \ruby also shows strong Balmer and Paschen emission lines, and a blue SED shape in the rest-frame near-infrared \citep{Setton2025}. Similar to A2744-QSO1, but unlike the most luminous LRDs, \ruby does not have strong \Oiii$_{\lambda5008}$ emission or any other strong metal lines (further quantified in Section~\ref{sec:elines}).

\subsection{Balmer break}

Most strikingly, \ruby stands out for its extremely strong Balmer break in comparison to the four literature sources. 
We quantify this in Figure~\ref{fig:BB_zspec}, where we compile a large sample of LRDs \citep{Furtak2024,Labbe2024,Setton2024,Wang2024b} and massive quiescent galaxies at $z>3$ \citep{Barrufet2025,Carnall2024,deGraaff2024c,Glazebrook2024,Nanayakkara2024,Weibel2024b} with public JWST/NIRSpec spectroscopy obtained from the DJA. To measure the Balmer break strength we integrate the spectrum in two tophat filters with wavelength ranges $[3620,3720]\,\AA$ and $[4000,4100]\,\AA$ using \texttt{pyphot} \citep{zenodopyphot}, and compute the flux density ratio (in $f_\nu$)\footnote{This definition differs slightly from \citet{Wang2024a}, who used the same wavelength ranges, but computed the ratio of the median flux density in $f_\lambda$. However, the definition used here more closely matches the commonly measured Dn4000 index of \citet{Balogh1999}, and also allows for better estimation of the uncertainty.}. Error bars are estimated using 500 random Gaussian draws from the error spectrum. For reference, we also show the Balmer break strengths measured from stacks of star-forming galaxies by \citet{RobertsBorsani2024}. 

The Balmer break of \ruby ($6.9_{-1.5}^{+2.8}$) is $\gtrsim2$ times stronger than that of any high-redshift massive quiescent galaxy ($<3.1$), and all high-redshift LRDs with Balmer breaks published to date (the largest break strength being $\approx3.56$ for A2744-45924 of \citealt{Labbe2024}). The only source (\mombh; \citealt{Naidu2025}) that matches the break strength of \ruby is a LRD that was discovered at approximately the same time in JWST Cycle 3 program GO-5224 (PIs: Oesch \& Naidu), and is discussed in detail in Section~\ref{sec:cloudy}.

\subsection{Emission lines}\label{sec:elines}

\begin{figure}
    \centering
    \includegraphics[width=\linewidth]{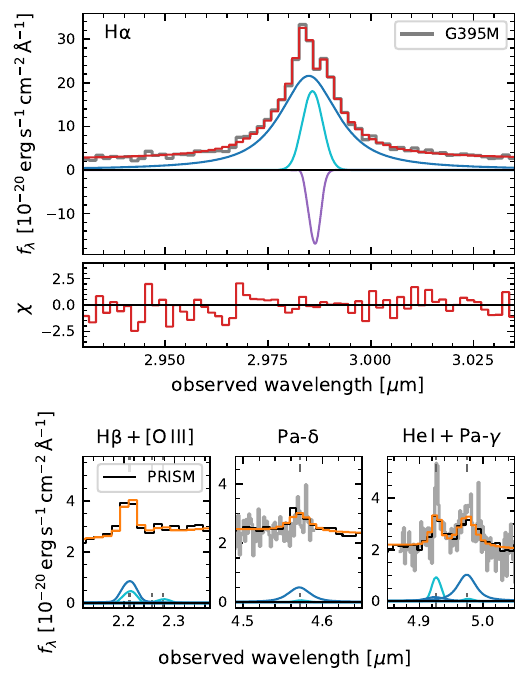}
    \caption{Kinematic emission line decomposition from simultaneous fitting to PRISM (black) and G395M (grey) spectra. Top: The \Ha emission and absorption complex is well-described by a broad Lorentzian profile (blue; $\rm FWHM\sim1400\,\kms$), with weaker narrow emission (cyan) as well as redshifted absorption (purple). The total G395M model and residuals are shown in red. Bottom: Zoom-ins of other strong H and He features, also revealing a non-detection of the \Oiii doublet (see Table~\ref{tab:elines}). Although (where possible) both dispersers were used in the fitting, in the bottom panels we show only the PRISM model components and total model (orange).}
    \label{fig:elines}
\end{figure}

The spectra reveal a suite of emission lines, primarily Balmer, Paschen and \Hei lines. The G395M spectrum covers the rest-frame wavelength range of $0.64-1.14\,\micron$ ($2.9-5.2\,\micron$ observed; with the chip gap falling at $4.6-4.8\,\micron$), and therefore kinematically resolves the \Ha line as well as various Pa features. Figure~\ref{fig:elines} shows zoomed in wavelength ranges of the strongest emission lines of both the PRISM and G395M spectra. 

The \Ha emission shows a complex, broad and asymmetric profile. We attribute this asymmetry to a redshifted Balmer absorption line, a feature that has been observed in several other LRDs and likely originates from absorbing dense gas clouds along the line of sight \citep[e.g.][]{Matthee2024,Juodzbalis2024,DEugenio2025}. It is likely that this absorption is also present in other lines (e.g. \Hei or \Hb; \citealt{Wang2024a,Ji2025}), but at the current signal-to-noise of the data ($S/N$) and available spectral resolution we cannot establish whether this is the case. 

We therefore begin by focusing on the high $S/N$ \Ha line complex in the G395M spectrum, using the Bayesian emission line fitting software {\texttt{unite}} described in {\citet{Hviding2025}}. Briefly, this builds on our prior fitting methods (as in \citealt{Wang2024a,Wang2024b,deGraaff2024d}) and therefore robustly accounts for the undersampling of the JWST/NIRSpec line spread function (LSF) through integration of the emission line profiles. The error spectra are automatically rescaled to match the measured scatter in the continuum blue- and redward of the emission line. Moreover, our fitting accounts for calibration uncertainty in the NIRSpec LSF: we assume the LSF of an idealised point source from \citet{deGraaff2024a}, but introduce a nuisance parameter $f_{\rm LSF}$ by which the dispersion is multiplied to model both the systematic and measurement uncertainty in the LSF; the prior for this nuisance parameter is a truncated Gaussian centred at $f_{\rm LSF}=1.2$ and dispersion of $0.1$, with minimum and maximum values of $[0.9,1.5]$. As in Section~\ref{sec:imaging}, the probabilistic model is implemented in \texttt{numpyro} using the No U-Turn Sampler to generate posterior samples.

Our fiducial model for the \Ha complex, presented in Figure~\ref{fig:elines} and Table~\ref{tab:elines}, assumes a broad Lorentzian and narrow Gaussian emission component, a narrow Gaussian absorption component, and a linear continuum. The redshifts of all components are allowed to vary; we use uniform priors for the line widths and fluxes. We have tested a range of model variations, which we do not show but describe here, performing model comparison to the fiducial model using the Widely Applicable Information Criterion \citep[WAIC;][although the Bayesian Information Criterion yields indentical conclusions]{Watanabe:WAIC}. First, we find that a broad Lorentzian component provides a significantly better fit over a broad Gaussian component\footnote{This difference in modelling likely also largely explains the difference between the broad \Ha FWHM obtained here and the larger FWHM measured by \citet{Taylor2024} for \ruby ($\rm FWHM\sim2300\,\kms$), who used a two-component Gaussian model. } ($\rm\Delta WAIC=39.8$), as was also found for the broad \Ha line of A2744-45924 \citep{Labbe2024}. Second, although the flux of the absorption component is poorly constrained in our fiducial model, the model with an absorption component is formally strongly preferred over one without ($\rm\Delta WAIC=27.6$). Deep, high-resolution observations would be needed to robustly constrain the kinematics and equivalent width of the absorber. Finally, we have tried to include the \Nii doublet in the fit (with fixed flux ratio of 1:2.95 for the doublet). This fit is weakly disfavoured compared to the fiducial model ($\rm\Delta WAIC=3.76$), and the flux of the \Nii doublet is consistent with zero and all other parameters are consistent with the values presented in Table~\ref{tab:elines}.

Next, we perform simultaneous fitting to the PRISM and G395M spectra to obtain emission line properties for a broader suite of lines. The key benefit of this approach is that it leverages the $S/N$ of both spectra. We introduce two new parameters in order to enable simultaneous fitting of multiple dispersers, which are necessary due to systematic uncertainties in the flux and wavelength calibration. Specifically, we fit for a flux offset between the G395M and PRISM dispersers as well as a detector pixel offset, and set Gaussian priors on these parameters based on the population average offsets from a large number of spectra \citep[for details on calibration issues and typical offsets, see][]{deGraaff2024d}. We use the following setup for the model: we assume a Lorentzian broad component, as well as a Gaussian narrow emission component for the \Hb, \Ha, \Pad, \Pag, \Hei$_{\lambda 10830}$ and \Oi$_{\lambda 8446}$ lines. For \Ha we also add a Gaussian absorption component. We further model the \Oiii$_{\lambda\lambda 4960,5008}$ emission lines with only a narrow Gaussian component, fixing the flux ratio of the doublet to 1:2.98. Due to the modest $S/N$ for most emission lines, all narrow and broad line velocity widths are tied together; the broad and narrow line redshifts are allowed to deviate. We further assume a linear continuum in the region around each emission line ($\pm 15000 \kms$).

The line fluxes from this joint fit are presented in Table~\ref{tab:elines}, and best-fit models are shown for the G395M (red) and PRISM (orange) spectra in Figure~\ref{fig:elines}. The narrow and broad component redshifts and velocity dispersions of this joint fit are consistent with our previous fit to the \Ha line alone, and we therefore do not list these separately. The \Oiii and \Oi emission lines are not significantly detected, which may imply a low metallicity or high gas density (see Section~\ref{sec:cloudy}), and we therefore provide the 95th percentiles as upper limits. The total \Hb emission is detected at the $\sim10\sigma$ level, but its broad and narrow components are individually poorly constrained due to the low spectral resolution of the PRISM at $\sim2\,\micron$ ($R\sim50$).

\begin{table}
\caption{Emission line properties. }
\centering
 \begin{tabular}{lll}\hline\hline
  \multicolumn{3}{c}{Line fluxes}\\\hline
  Line & Flux (narrow) & Flux (broad)\\
     & ($\rm 10^{-18}\, erg\,s^{-1}\,cm^{-2}$) & ($\rm 10^{-18}\, erg\,s^{-1}\,cm^{-2}$) \\\hline
  \Hb &   $1.7_{-1.2}^{+1.7}$ & $3.8_{-2.0}^{+1.6}$ \\  
  \Oiii$_{\lambda\lambda 4960,5008}$ &  $<1.5$ &  \\  [2pt]
  \Ha (emission) &  $13.0_{-3.0}^{+4.8}$ & $61.6_{-2.0}^{+1.8}$ \\ [2pt] 
  \Ha (absorption) &  $-7.2_{-5.1}^{+2.2}$ &  \\  [2pt]
  \Oi$_{\lambda 8446}$ &  $<0.5$ & $<0.9$ \\  [2pt]
  \Pad &   $0.13_{-0.09}^{+0.16}$ & $2.4_{-0.4}^{+0.4}$ \\  [2pt]
  \Pag &   $0.15_{-0.11}^{+0.18}$ & $5.2_{-0.5}^{+0.4}$ \\  [2pt]
  \Hei$_{\lambda 10830}$ &  $1.8_{-0.3}^{+0.3}$ & $0.9_{-0.5}^{+0.5}$ \\  [2pt]\hline\hline
  \multicolumn{3}{c}{\Ha kinematic properties}\\\hline
   $z_{\rm narrow}$ & $3.5484_{-0.0004}^{+0.0004}$ & \\[2pt]
   $z_{\rm broad}$ & $3.5470_{-0.0003}^{+0.0003}$ & \\[2pt]
   $\Delta v_{\rm absorb}$ &  $49^{+26}_{-21}\,\kms$ & \\[2pt]  
   $\rm FWHM_{\rm narrow}$ & $478^{+95}_{-93}\,\kms$  &  \\[2pt]
   $\rm FWHM_{\rm broad}$ &  $1533^{+110}_{-80}\,\kms$ & \\[2pt]
   $\rm FWHM_{\rm absorb}$ & $133^{+95}_{-92}\,\kms$  & \\[2pt]  \hline
 \end{tabular}
\label{tab:elines}
\end{table}

\section{Morphology}\label{sec:morph}

We previously (Section~\ref{sec:imaging}) established that \ruby has a compact morphology. However, it is challenging to quantify robustly whether a LRD is extremely compact or truly unresolved, as many high-redshift LRDs are very faint in the short wavelength NIRCam filters that have the highest spatial resolution. Only for few objects has it been possible to obtain robust upper limits on the size based on such rest-UV imaging \citep[e.g.,][upper limit of $<30\,$pc]{Furtak2024}. Others have typically either assumed conservative upper limits on the size \citep[e.g. the half width at half maximum at $4\,\micron$ of $\lesssim 400\,$pc,][]{Leung2024}, or focused on S\'ersic profile fitting in the long wavelength NIRCam filters \citep[e.g.][resulting in size upper limits of $\lesssim100-200\,$pc]{Baggen2023,Baggen2024,Akins2024}. Because of its (comparatively) lower redshift, the target of this paper is fortunately bright also in the reddest short-wavelength NIRCam filter (F200W, probing rest-frame $0.44\,\micron$) for which the PSF $\rm FWHM\approx0.066\arcsec$.

The second uncertainty in modelling compact morphologies is the PSF model itself, which is typically constructed empirically from a stack of stars selected across the entire mosaic. Therefore, it represents an average PSF that may deviate in detail from the true local PSF due to small geometric distortions. Here we use the empirical PSFs constructed using the methods described in \citet{Weibel2024}, with stars selected in the magnitude range $\rm F200W\sim19-24.5$. 

\begin{figure}
    \centering
    \includegraphics[width=\linewidth]{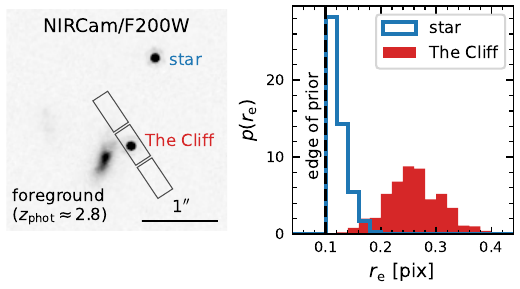}
    \caption{Left: NIRCam/F200W image ($0.02\,\arcsec\,{\rm pix}^{-1}$) of \ruby, the nearby foreground source at a lower redshift of $\zphot\approx2.8$, and a star of approximately equal magnitude (at a distance of $\sim1\arcsec$). All sources are modelled as single S\'ersic profiles and fit simultaneously to account for blending. Right: The posterior distribution of the effective radius of the star is clustered at the edge of the prior, as may be expected for an unresolved point source, while that of the LRD converges to a small value of $\re\approx0.26\,\pix\approx 40\,$pc. }
    \label{fig:sersic}
\end{figure}

Serendipitously, a star of $\rm F200W=25.1$ lies $\sim 1\arcsec$ away from \ruby (see Figure~\ref{fig:sersic}). This star is too faint to have been included in our empirical PSF model, but it is approximately equally bright as the primary target (for which $\rm F200W=25.3$). In other words, the nearby star provides an ideal test of our empirical PSF model and S\'ersic profile fitting.

We perform simultaneous fitting with \texttt{pysersic} as already described in Section~\ref{sec:imaging}, modelling the foreground source, star, and LRD with single S\'ersic profiles, but now for the F200W mosaic with a pixel scale of $0.02\arcsec$ {(see Appendix~\ref{sec:apdx_photometry} for the MAP model and residual image)}. This time, the prior on the S\'ersic index ($[0.65,6.0]$) and effective radius ($[0.1,100]\,\pix$) are allowed to be broader. We find that if we use a lower bound of $\re=0.5\,\pix$, the posterior distributions of the star and LRD are identical for the size, axis ratio and S\'ersic index. However, when we lower this bound to $0.1\,\pix$ these posteriors begin to deviate. The axis ratios and S\'ersic indices are consistent within the uncertainties (Table~\ref{tab:morph}), but the sizes differ significantly. In Figure~\ref{fig:sersic} we show that the posterior of the effective radius piles up at the boundary of the prior for the star, which is expected for an unresolved point source, and therefore can be interpreted as a success of our empirical PSF model. On the other hand, the fit to the LRD converges to a very small value of $\re\sim0.26\,\pix$, corresponding to a physical major axis size of $\sim 40\,$pc. 

The LRD therefore could formally be considered to be marginally resolved. However, it is also possible that a bright point source dominates over a faint (more extended) host galaxy, leading to a slightly resolved size when fitting a single S\'ersic component. We further caution that the nearby foreground source appears to have diffuse extended emission, and its morphology deviates from a simple, symmetric S\'ersic profile. This residual flux may therefore bias the inferred size of the LRD to higher values. Nevertheless, our S\'ersic profile fitting does provide a stringent upper limit on the size of $<51\,$pc (95th percentile; or $<58\,$pc, 99.7th percentile), which is substantially lower than most upper limits for (unlensed) LRDs in current literature.

\begin{table}
\caption{Morphological properties.}
%\begin{center}
\centering
 \begin{tabular}{lcccc}\hline\hline
  Source & $\re$ (pix) & $\re$ (pc) & axis ratio & S\'ersic index \\ \hline
   %  & (pix) & &  \\\hline
  star &  $0.117_{-0.011}^{+0.021}$ &  $17.5_{-1.7}^{+3.2}$ & $0.41_{-0.22}^{+0.30}$ & $0.92_{-0.19}^{+0.32}$  \\[4pt]  
  LRD &  $0.259_{-0.046}^{+0.050}$ &  $38.6_{-6.9}^{+7.4}$ & $0.47_{-0.26}^{+0.29}$ & $1.17_{-0.30}^{+0.42}$ \\[2pt]  \hline
 \end{tabular}
\label{tab:morph}
%\end{center}
\end{table}

\section{Stellar population modelling}\label{sec:sed_models}

We now turn to the physical interpretation of the SED, and ask whether the observed Balmer break and spectral shape could be explained by conventional stellar population models. A range of models and codes have been used in recent literature to fit the SEDs of LRDs, with varying degrees of success (as described in Section~\ref{sec:intro}). Here, we focus only on tools that are able to fit spectroscopic data and on models that are able to produce a (sharp) Balmer break. Specifically, we apply the modelling of \citet{Wang2024a,Wang2024b} and \citet{Labbe2024}, but do not consider their AGN-only or AGN-dominant models, as those were already shown to fail to produce strong enough Balmer breaks for LRDs that are less extreme than \ruby.

\subsection{\prospector models}\label{sec:prospector}

Following the methodology described in \citet{Wang2024a,Wang2024b}, we jointly fit the PRISM spectrum and NIRCam+MIRI photometry with the Bayesian modelling software \prospector \citep{Leja2017,Johnson2021}, using the dynamic nested sampling code \texttt{dynesty} \citep{Speagle2020}. Model spectra are redshifted using the best-fit redshift of Section~\ref{sec:elines} and convolved to the NIRSpec/PRISM resolution using the model LSF of an idealised point source of \citet{deGraaff2024a}. To account for small mismatches between the flux calibration of the spectrum and photometry, we allow for a constant rescaling (i.e. a zeroth-order polynomial calibration vector) between the spectrum and the photometry. We find that higher order polynomials ($n\geq2$) lead to calibration vectors that are significantly model-dependent in shape and amplitude, indicating that this vector corrects for the mismatch between model and data rather than any real flux offset between the spectrum and photometry. Emission lines are not interpreted physically in the modelling, i.e., we fit broad Gaussians for the \Hb, \Ha, \Pad, \Pag and \Hei lines with flux ratios that are allowed to vary arbitrarily. 

\begin{figure*}
    \centering
    \includegraphics[width=0.97\linewidth]{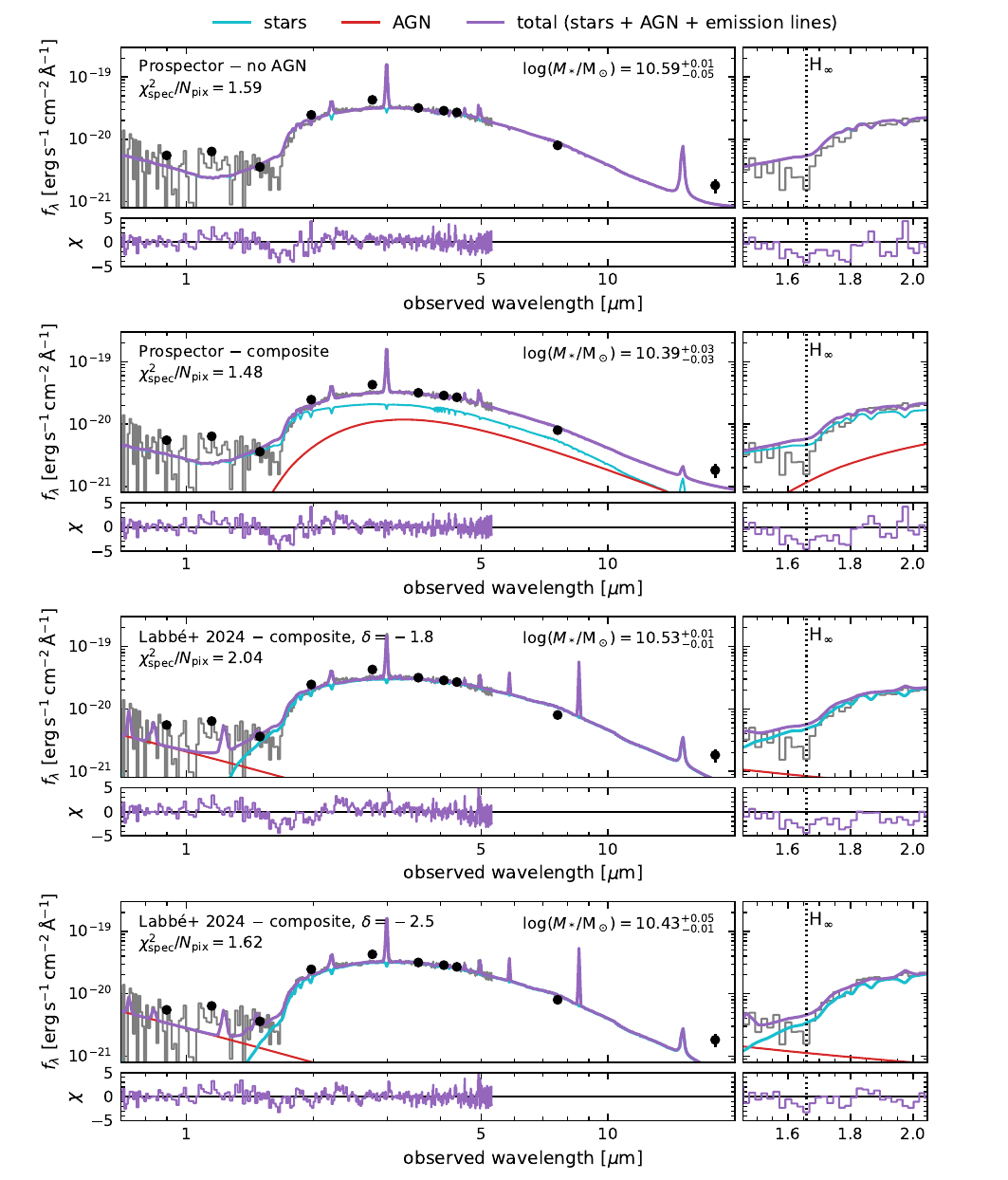}
    \caption{Best-fit SED models and residuals (with respect to the PRISM spectrum) for four model variations, constructed as different mixtures of (dust-reddened) stellar population and power-law AGN model components. These AGN models do not consider the possible reddening by absorbing dense gas, which we explore in Section~\ref{sec:cloudy}. {Right-hand panels show zoom-ins of the region around the Balmer break.}  
    From top to bottom: the fiducial `galaxy-only' model from \prospector; a galaxy + AGN model that maximises the stellar contribution, fit with \prospector (the `maximal $M_*$' model of \citealt{Wang2024b}); the galaxy + AGN model following \citet{Labbe2024}, but fitting only Hydrogen emission lines instead of a forest of metal lines; the galaxy + AGN model of \citet{Labbe2024}, but with an even steeper dust law (see Section~\ref{sec:dust}). All four models favour a massive post-starburst solution, with a very steep dust attenuation law and high optical depth. However, even with such extreme dust, none of these models can produce the strong Balmer break and shape of the rest-frame optical SED, as is evident from the systematic (and significant) features in the residuals blue- and redward of the Balmer break. }
    \label{fig:seds} 
\end{figure*}

\begin{table*}
\caption{Stellar population fitting results. }
%\begin{center}
\centering
 \begin{tabular}{lllll}\hline\hline
  Parameter & \prospector  & \prospector  & Labb\'e et al. & Labb\'e et al.  \\ 
    & (galaxy-only / no AGN) & (galaxy + AGN) & (galaxy + AGN) & (galaxy + AGN; extreme dust) \\\hline
   %  & (pix) & &  \\\hline
  $\log(M_*/\Msun)$ &  $10.586_{-0.048}^{+0.009}$ &  $10.39_{-0.03}^{+0.03}$ & $10.531_{-0.008}^{+0.007}$ & $10.434_{-0.049}^{+0.014}$ \\[2pt]  
  age [Gyr] &  $0.90_{-0.42}^{+0.11}$ &  $0.69_{-0.11}^{+0.11}$ & $0.65_{-0.04}^{+0.04}$ & $0.66_{-0.19}^{+0.10}$ \\[2pt]  
  $\log(Z/Z_\odot)$ &  $-1.95_{-0.02}^{+0.14}$ &  $-1.84_{-0.09}^{+0.08}$ & $-0.986_{-0.010}^{+0.023}$ & $-0.83_{-0.12}^{+0.3}$ \\[2pt]  
  SFR $[\Msun\,\rm  yr^{-1}]$ &  $7.3_{-4.0}^{+6.0}$ &  $0.43_{-0.40}^{+0.95}$ &  \\[2pt]
  $\log(\tau/{\rm yr})$ & & & $6.9_{-0.3}^{+0.3}$ &  $7.0_{-0.4}^{+0.3}$   \\[2pt]  
  $A_{\rm V}$ (stars) &  $1.63_{-0.08}^{+0.13}$ &  $1.43_{-0.05}^{+0.05}$ & $1.34_{-0.02}^{+0.02}$ & $1.06_{-0.03}^{+0.04}$  \\[2pt]  
  dust index &  $-0.991_{-0.007}^{+0.013}$ &  $-0.97_{-0.02}^{+0.03}$ & -1.8 &  $-2.47_{-0.02}^{+0.04}$ \\[2pt]  \hline
 \end{tabular}
\tablefoot{The dust index corresponds the power-law slope for the \citet{Labbe2024} models. For the \prospector models it instead corresponds to the \emph{corrective} power-law slope to the Calzetti dust law (i.e. $\delta$ in the parametrisation of \citealt{Noll2009}). }
\label{tab:sed_fitting}
%\end{center}
\end{table*}

\subsubsection{Galaxy-only model}

We begin by fitting a model that includes a stellar population component as well as emission lines, but no AGN component: this is the galaxy-only model of \citet{Wang2024a}, and uses the model setup and informative priors of the \prospector-$\beta$ model described in \citet{Wang2024:pbeta}. Briefly, this uses the Flexible Stellar Population Synthesis models \citep[FSPS;][]{Conroy2010}, with MIST stellar isochrones \citep{Choi2016,Dotter2016} and the MILES stellar library \citep{Sanchez-Blazquez2006}, and assumes a \citet{Chabrier2003} initial mass function (IMF; the possibility of IMF variations is discussed in Section~\ref{sec:imf}). The star formation history follows a modified version of the non-parametric \prospector-$\alpha$ model of \citet{Leja2017} with 7 logarithmically-spaced times bins that are optimised for the high-redshift Universe. Dust attenuation follows a two-component dust screen model \citep{Charlot2000} with the flexible attenuation curve of \citet{Noll2009}, and dust emission is included using the models of \citet{DraineLi2007}. Moreover, a fraction of the stellar emission ($f_{\rm nodust}$) is allowed to remain outside this dust screen. {A complete description of the model parameters and adopted priors can be found in \citet[][Table 1]{Wang2024:pbeta} and \citet[][Table 2]{Wang2024a}.}

The best-fit galaxy-only model implies a massive, post-starburst galaxy (Table~\ref{tab:sed_fitting}) and a dust attenuation law that is extremely steep, with a dust index that reaches the edge of the prior {(minimum $\rm dust\ index=-1.0$)}. Only the combination of an evolved stellar population, with a
low recent star formation rate (SFR), extremely low metallicity yet strong dust attenuation can produce a strong Balmer break as well as faint UV emission. However, Figure~\ref{fig:seds} (top panel) shows that this is still not sufficient to explain the observed spectrum, as significant residuals remain blue- and redward of the Balmer break. Because of the steep dust law, the model also slightly overpredicts the emission in the rest-frame near-IR spectrum and photometry (observed wavelength range of $4-8\,\micron$).

\subsubsection{Galaxy + AGN model}

Next, we add an AGN model to perform a composite fit, following the procedure described in \citet{Wang2024a}. The intrinsic SED of this AGN accretion disc model is constructed from piece-wise power laws with indices fixed to the best-fit values of \citet{Temple2021}. This SED is reddened by the same dust law that is applied to the stellar population, and additionally reddened by a second power law extinction curve of which the slope and normalisation are allowed to vary. We further include a model for the hot dust emission of an AGN torus, using the CLUMPY torus model \citep{Nenkova2008b} as implemented in FSPS \citep{Conroy2009,Leja2018}. The total AGN model therefore introduces five new free parameters: three of these parametrise the additional dust law (2) and optical depth of the torus (1), and the remaining two quantify the flux ratio between the galaxy and AGN at rest-frame $5500\,\AA$ ($f_{\rm AGN}$) and ratio of the AGN mid-IR luminosity and the galaxy bolometric luminosity  ($f_{\rm AGN, torus}$).

These fits can be tuned in order to down-weight or up-weight the galaxy versus AGN components, by setting different priors on $f_{\rm AGN}$ and the total mass formed. \citet{Wang2024b} hence proposed three different model flavours, resulting in models with a minimal, medium and maximal stellar contribution, with the `maximal $M_*$' model yielding the greatest Balmer break strengths. {The priors used for the latter model are as described in \citet[][Table 2]{Wang2024a}, with the exception of $f_{\rm AGN}$ and the total mass formed, for which we assume a lognormal distribution with $\mu=-3$ and $\sigma=1$ and a log-uniform distribution respectively.}

Here, we present only this `maximal $M_*$' model in Table~\ref{tab:sed_fitting} and Figure~\ref{fig:seds}, although we have also tested the other variants, finding significantly worse fits. The galaxy component dominates at the Balmer break and in the rest-optical, and the AGN only contributes substantially (although still sub-dominant) in the rest near-IR. The dust law required is similarly steep as for the galaxy-only model. Including the AGN component provides a marginally better fit in the rest-frame optical ($\sim 2-4\,\micron$) compared to the galaxy-only model, but still cannot fit the Balmer break.

\subsection{Labb\'e et al. composite model}\label{sec:labbe_models}

Recently, \citet{Labbe2024} developed an independent full spectral fitting approach tailored to a luminous LRD with a strong Balmer break (the second strongest break in Figure~\ref{fig:BB_zspec}). This fitting shares some of the features of the \prospector modelling, such as its Bayesian approach (performing nested sampling with \texttt{pymultinest}; \citealt{Buchner2014}), and the use of FSPS models with a Chabrier IMF. It differs in other aspects, such as the fact that only the spectrum (i.e. no photometry) is used for fitting. Moreover, the star formation history is parametrised as a delayed $\tau$ model, {and the dust model is a power law with slope $\delta$ and variable normalisation. We use the exact same priors as described in detail in Table~6 of \citet{Labbe2024}, with the exception of the dust law, for which we fix $\delta=-1.8$ (the lower bound of the prior range considered in their work).} We will assess the choice of this dust law in Section~\ref{sec:dust}.

The most important differences are in the AGN component: this is modelled as a single power law continuum with variable slope $\beta$, and combined with a large suite of emission lines, such as the Balmer and Paschen series (with relative intrinsic fluxes assuming case B recombination), Fe\,{\sc ii} emission, and other lines commonly observed in AGN (e.g. \Hei, He\,{\sc ii}, Ne\,{\sc v}, O\,{\sc i}). Depending on line species, the emission is broadened according to the \Ha line width or the narrow \Oiii emission. The AGN emission is reddened by a power law of the same slope as the dust law applied to the stellar population, but with a different normalisation. 

Three different models were constructed with these components. The first model is a two-component blue+red AGN (where the blue AGN component is an unattenuated fraction of the red component), and the second a two-component galaxy model without any AGN. Neither model was shown to successfully produce a strong Balmer break. We therefore focus only on the third model here, which combines the two-component AGN model with a single stellar component. We make one modification to this model: because we do not observe any significant metal lines in \ruby (Section~\ref{sec:elines}), we only include broad Hydrogen lines in this fit (neglecting \Hei).

The resulting fit is shown in the third panel of Figure~\ref{fig:seds}. The best-fit AGN model is clearly substantially bluer than in the \prospector galaxy+AGN model. Otherwise, the fit looks remarkably similar to the two \prospector models, yielding a massive post-starburst stellar population. As before, the combination of an older stellar population and a steep attenuation law is required in order to produce a strong Balmer break. The model fails in the same way as the \prospector models, as it does not capture the magnitude of the observed Balmer break, resulting in a systematic over- and under-prediction of the flux blue- and redward of the break, respectively. The model also slightly overpredicts the rest near-infrared, although the MIRI photometry was not included in the fit.

\section{Discussion}\label{sec:discussion}

We have shown that \ruby has a highly unusual spectrum with a dramatic Balmer break that proves impossible to reproduce with typical evolved stellar populations. While massive post-starburst galaxies typically show the strongest Balmer breaks, as their SEDs are dominated by A stars, the exceptionally strong Balmer break of \ruby vastly exceeds measurements of post-starburst galaxies at $z\gtrsim3$. Only a steep dust attenuation curve can substantially increase this break strength. Therefore, if the rest-optical luminosity were dominated by starlight, the galaxy would have to be massive, strongly dust-attenuated, and abruptly ceased star formation in the last few hundred Myr.

In what follows, we will test each of these characteristics. We first consider whether a dust law necessary to match the observed spectrum can be found, and whether it is realistic (Section~\ref{sec:dust}). Next, we ask whether a high stellar mass within a very compact galaxy is plausible, by exploring the consequences of high stellar mass densities on the galaxy dynamics (Section~\ref{sec:collisions}). We then consider our assumption of a universal IMF and the contribution of A stars to the spectrum (Section~\ref{sec:imf}), which underpins the SED modelling of Section~\ref{sec:sed_models}. Finally, we discuss possible alternative models for \ruby, in particular focusing on a new class of AGN-dominated models in Section~\ref{sec:cloudy}.

\subsection{Peculiar dust?}\label{sec:dust}

\begin{figure}
    \centering
    \includegraphics[width=\linewidth]{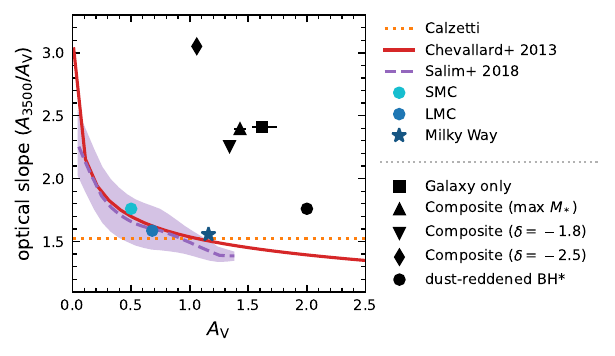}
    \caption{Optical dust attenuation \emph{slope} across the Balmer break ($A_{3500}/A_{\rm V}$) vs. the optical attenuation $A_{\rm V}$. Radiative transfer models predict a strong correlation between the slope and optical depth \citep{Chevallard2013}, which closely matches observed dust curves \citep{Gordon2003,Gordon2009,Salim2018}. The various black symbols denote optical slopes and depths implied by the SED models of \ruby { from Section~\ref{sec:sed_models} (medians are shown with 16-84 percentiles, although these error bars are in most cases smaller than the markers)}, which deviate strongly from this trend. {The dust-reddened BH* model is discussed in Section~\ref{sec:cloudy}.}}
    \label{fig:dust}
\end{figure}

The primary point here to be stressed about the SED models in Section~\ref{sec:sed_models} is that they imply an extreme, arguably implausible, combination of steep optical slopes \emph{and} high optical depths ($A_{\rm V}\sim1.4-1.7$) for the dust attenuation. With power-law slopes $\delta\sim-1.8$, the implied UV-optical slopes $A_{1500}/A_{\rm V}\sim 10$, a factor $\sim 2$ steeper than any individual sightline in the SMC \citep{Gordon2003}. Most important in the context of this work is the reddening across the Balmer break, and instead of the UV-optical slope we therefore measure the \emph{optical} slope of the dust law as the ratio of the attenuation at $3500\,\AA$ and $5500\,\AA$ ($A_{3500}/A_{\rm V}$) in Figure~\ref{fig:dust}. Clearly, both the galaxy-only model and the two composite AGN+galaxy models discussed in Section~\ref{sec:sed_models} have considerably steeper optical slopes ($30-60\%$) than the \citet{Calzetti2000} dust law or the average Milky Way, SMC and LMC dust curves \citep[][measured using the \texttt{dust\_extinction} package of \citealt{Gordon2024}]{Gordon2003,Gordon2009}. 
%Moreover, the attenuation is extraordinarily high given the steep slope, with $A_{\rm V}\sim1.4-1.7$. 

Observations of large samples of galaxies \citep[e.g.][]{Salim2018}, radiative transfer models \citep[e.g.][and references therein]{Chevallard2013}, and hydrodynamical simulations \citep[at fixed grain composition, e.g.][]{Narayanan2018} all show that such steep dust curves only occur in concert with low optical depths ($A_{\rm V}\lesssim0.5$), as the result of the star-dust geometry \citep[for a review see][]{Salim2020}. Conversely, high optical depths are observed and modelled only in systems with greyer attenuation curves. The steep slopes and high optical depths of our SED models therefore lie outside the parameter space of observations and models so far (Figure~\ref{fig:dust}), raising the question whether a suitable star-dust geometry or grain composition even exists that could produce such an attenuation curve. 

Considering dust laws that are even steeper, yet optically thick at rest $\sim 4000\,\AA$, therefore appears physically implausible. Nevertheless, we ask if allowing for such a dust law would yield a better fit to the spectrum. We rerun our SED fitting for the \citet{Labbe2024} composite AGN+galaxy model with the power-law index as a free parameter, extending the lower bound of the prior on the dust law slope to $\delta=-2.5$. The UV-optical (optical) slope of this new bound is a factor $\sim 5$ ($\sim 2$) steeper than seen in SMC sightlines, and difficult to produce also in hydrodynamical simulations that investigate the dust-star geometry \citep{Narayanan2018,Salim2020}. The posterior distribution for this model pushes against the boundary of $\delta=-2.5$ and with high $A_{\rm V}\sim1.2$, a factor $\sim10$ greater optical depth than expected for such a steep curve (Figure~\ref{fig:dust}). %the extra differential reddening in the near-infrared is compensated for by a slightly older age of the stellar population. 
The resulting model is shown in the bottom panel of Figure~\ref{fig:seds}, and provides the formally best fit in the Balmer break region out of all four models presented. 

However, significant residuals around the Balmer break remain, even with this extremely steep (and likely unrealistic) dust law. Previous studies of LRDs with strong Balmer breaks had already demonstrated that steep dust laws and high optical depths are necessary to fit the SED \citep{Ma2024,Labbe2024,Wang2024b}. But, the LRDs presented in these papers had Balmer breaks a factor $\sim2$ less strong than that of \ruby, which in principle could be produced with the dust laws considered in Section~\ref{sec:sed_models}. 

Because the Balmer break of \ruby is so exceptionally strong, we can now robustly conclude that massive stellar population models cannot fit the spectrum, even when considering extreme dust laws.

\subsection{Kinematics of ultra-dense stellar systems}\label{sec:collisions}

\begin{figure*}
    \centering
    \includegraphics[width=0.85\linewidth]{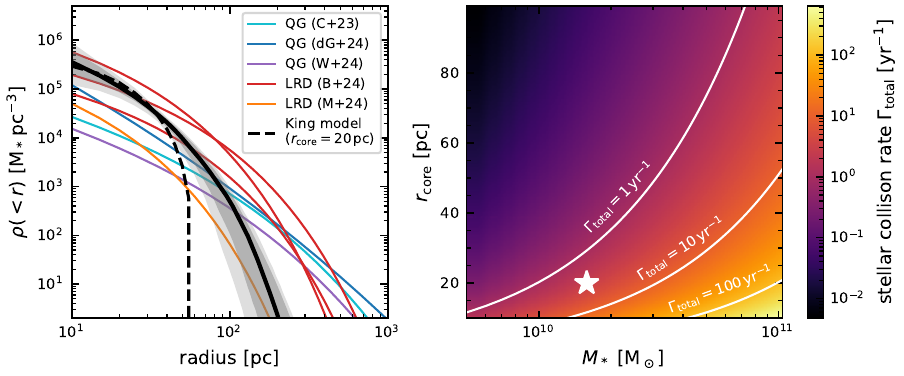}
    \caption{Left: Stellar mass density profiles of \ruby (black solid; grey shaded areas indicate 16-84 {and 5-95} percentiles), a selection of compact massive quiescent galaxies at $z>4$ \citep[cyan, blue, and purple lines;][]{Carnall2023,deGraaff2024c,Weibel2024b} and LRDs with strong Balmer breaks from the literature \citep[orange and red lines;][]{Baggen2024,Ma2024}. The stellar mass densities of LRDs, under the assumption that the rest-optical is dominated by stars, exceed the densities in the cores of massive quiescent galaxies by an order of magnitude. Right: Stellar collision rate $\Gamma_{\rm tot}$ as a function of the core radius (as defined in the King model) and stellar mass of the system, for a $\sim 500\,$Myr old stellar population. The inner $\lesssim50\,$pc of \ruby can be well approximated with a King model of $r_{\rm core}\sim20\,$pc; {the location of the star corresponds to the dashed line shown in the left panel}. If the stellar mass is as high as implied by the SED modelling of Section~\ref{sec:sed_models}, we expect stellar collisions to occur at a frequency of $\sim 5\,{\rm yr}^{-1}$ (a factor $\sim10^5$ greater than for the Milky Way's nuclear star cluster). }
    \label{fig:mass_profile}
\end{figure*}

Before we assess a key assumption in our SED modelling -- the shape of the IMF -- we consider the implications that the high stellar masses of our explored SED fits would have on the galaxy's dynamical properties. We found in Section~\ref{sec:morph} that \ruby appears very compact  ($\re\sim40\,$pc) and would have to have a stellar masses of $\log(M_*/\Msun)\sim10^{10.4-10.6}$ (Table~\ref{tab:sed_fitting}), implying an extremely high mass density: the stellar mass surface density within the (circularised) effective radius $\Sigma_*(<\rec)\sim 6-9\times 10^{6}\,\Msun\,{\rm pc}^{-2}$ exceeds the maximum value from observations and models of star clusters and galactic nuclei by over an order of magnitude \citep[$\Sigma_{*,\rm max}\sim 3\times 10^5\,\Msun\,{\rm pc}^{-2}$;][]{Hopkins2010,Grudic2019}.

We deproject the S\'ersic profile and estimate the 3D mass density profile, $\rho$, following \citet{Bezanson2009}. This assumes spherical symmetry, and that the posterior distribution of Section~\ref{sec:morph} reflects an accurate measurement of the source morphology. {The left panel of Figure~\ref{fig:mass_profile} shows the median mass density profile (solid black line), with 16-84 (5-95) percentiles in dark (light) grey shading computed from 100 draws of the posteriors of the S\'ersic profile fit and a range in stellar masses that bracket the different models of Table~\ref{tab:sed_fitting}. We compare to the mass density profiles of massive quiescent galaxies at $z>4$ \citep[cyan, blue, and purple lines]{Carnall2023,deGraaff2024c,Weibel2024b}, and }also show the mass density profiles of LRDs with strong Balmer breaks, using stellar mass estimates that assume the rest-optical is dominated by stars \citep[orange line; from][]{Ma2024,Wang2024b} and size measurements (or upper limits) from \citet{Furtak2024} and \citet[][red lines]{Baggen2024}. \ruby is at least an order of magnitude denser than the cores of high-redshift massive quiescent galaxies, but matches the inferred (under the assumption that the optical is star-dominated) high stellar densities in some LRDs.

Such high stellar mass densities would imply that the stellar velocity dispersions are extremely high, with $\sigma_*\sim 10^3\,\kms$, as also discussed by \citet{Baggen2024}. \citet{Guia2024} proposed that such high densities would in turn result in a system that is dynamically unstable to gravitational collapse, and \citet{Bellovary2025} suggest that tidal disruption events from the resulting runaway collapse could explain the properties of LRDs. 

We consider either scenario unlikely, at least for \ruby. First, tidal disruption events that show broad Balmer emission lines typically have very blue SED shapes \citep[reviewed in][]{Gezari2021}, inconsistent with the observed SED of \ruby. Second, the crossing time of a collisionless stellar system $t_{\rm cross}= R/v\sim 10^{4-5}\,$yr for $R\sim10^{1-2}\,$pc and $v\sim 10^3\,\kms$. Assuming that the system comprises $N=10^{10}$ stars of equal mass, the timescale for dynamical relaxation \citep[Eq. 1.38 of][]{BinneyTremaine} is 
\begin{equation}
    t_{\rm relax} = \frac{0.1 N}{\ln N} t_{\rm cross} \sim 10^{2-3}\,\rm Gyr\,,
\end{equation}
and the core collapse timescale is approximately $t_{\rm cc} = 0.2\times t_{\rm relax}\sim 10^{1-2}\,{\rm Gyr}$ for such dense stellar systems \citep{Portegies2002}, thus much longer than the age of the Universe at $z=3.5$. 

However, in the above we have treated stars as point-like particles. If the density is very high, and the system contains a significant fraction of evolved stars, we may expect stellar collisions to occur. In the following, we quantify the rate of such collisions, and explore whether there are observable consequences. 

Generically, a local collision rate is defined via
\begin{equation}
    \Gamma_{\text{local}}(r) = \frac{1}{2}n(r)^2 A \langle v_{\text{rel}} \rangle,
\end{equation}
where $n(r)$ is the local number density [cm$^{-3}$], $A$ is the collision cross-section [cm$^2$] = $X_* (M/M_\odot) \pi R_\odot^2$, $X_*$ is the mean stellar disc area per unit mass in solar areas per solar mass, and $v_{\text{rel}}$ is the relative velocity between stars. Integrating over the whole system leads to a total collision rate of
$$\Gamma_{\text{total}} = \int_0^{r_t} \Gamma_{\text{local}}(r)4\pi r^2dr\,.$$

Taking a King model for such dense stellar systems, one gets 
$n(r) = n_0\exp(-\psi(r))$ for the density profile, $$M = 4\pi m_* \int_0^{r_t} r^2 n(r) dr = 4\pi m_* n_0 r_c^3 I(\psi_t)$$ 
for the total mass, where $I(\psi_t)$ is a dimensionless function of the truncation parameter and $m_*$ is the average stellar mass. The central velocity dispersion is related to the core radius: $\sigma_{v,0}^2 = 4\pi/9\,  G m_* n_0 r_{\rm core}^2.$
This eventually leads to a total collision rate of
\begin{equation}
    \Gamma_{\text{total}}(\text{yr}^{-1}) \approx 3.7\times10^{-7}\alpha X_*\left(\frac{M}{10^5M_\odot}\right)^2\left(\frac{r_{\rm core}}{1\text{ pc}}\right)^{-5/2},
\end{equation}
for $M$ in solar masses, $r_{\rm core}$ in parsecs, and $X_*$ in solar areas per solar mass. The constant $\alpha\approx 0.1$ arises from the King profile (for a King parameter $W_0\approx 7$).

Next, we need to calculate the cross section (to direct stellar collisions) per unit mass for a given stellar 
mass. For the vast velocities under consideration here, gravitational focussing does not dominate, and the cross section is simply geometric.
We consider Padova isochrones (with a Kroupa IMF) and integrate over the full range of masses $m$: $X_*\equiv \int A_*(m)\, p_{\rm Kroupa}(m)\, dm$, where $A_*(m)\equiv \pi R_*^2$ is the area of a star with mass $m$. For populations of several hundred Myr age we find values of $X\sim 5\, \pi R_\odot^2\, / \, M_\odot $. The right panel of Figure~\ref{fig:mass_profile} shows the total stellar collision rate as function of mass and core radius. 
For \ruby, a King model with $r_{\rm core}=20$~pc, $M=10^{10.3-10.4}M_\odot$ (i.e. approximately  half the stellar mass implied by the SED models) and $\alpha=-0.1$ provides a good description of the inner density profile (black dashed line); for $X=5$ this leads to $\Gamma_{\rm total}\sim4-6\,{\rm yr}^{-1}$. {Even for the lowest stellar mass found in Section~\ref{sec:sed_models} and upper limit on the size, we still find $\Gamma_{\rm total}=1\,{\rm yr}^{-1}$. }
Previously, such high-speed collisions have only been considered for the immediate vicinity of supermassive black holes \citep[creating velocities of $\gtrsim 1000\,\kms$;][]{Ryu2024, Hu2024}, where the collision rates are 
dramatically lower (e.g. $10^{-4.5}\,$yr$^{-1}$, for the Milky Way's nuclear cluster). 

What happens if the envelope of an evolved star is pierced in a random spot, e.g. at $R_*/\sqrt{2}$, by a solar-type main sequence star deserves, and quantitatively requires, hydrodynamical simulations that are beyond the current scope. High-speed stellar collisions have been simulated hydrodynamically in the literature \citep[e.g.][]{Ryu2024}, but not in the regime most likely here.  Instead, we can offer an order-of-magnitude toy scenario. The main sequence star, acting at $\sim 800\,\kms$ like a `bullet' piercing the evolved star's envelope, will create a bow shock that exceeds in extent the radius of the piercing star by a few times ($3-5\times$). The material then swept up within a cylinder of $5R_\odot$ at a projected centre distance of $40R_\odot$ in an evolved star of $R\approx 60R_\odot$ and $3M_\odot$ will sweep about $0.01M_\odot$ of the material. One can then envision that this material is heated to the virial temperature of the collision, of about $5\times 10^7$K for a collision at $800\,\kms$, corresponding to a thermal energy of $10^{47.5}\,\rm erg$ and a blackbody peak in the X-rays at 4.5\,keV.  How much of this energy will be radiated away, and how much will drive the expansion is unclear.
The thermal energy alone could afford a luminosity of $10^{42.5}\,\ergs$ for a week, which is shorter than the time between collisions and therefore might appear as X-ray outbursts. Although the expected luminosity is above the existing X-ray flux limits for \ruby (Section~\ref{sec:xray}), we cannot definitively rule out stellar collisions in this particular system due to the likely time variable nature of the X-ray emission.

Of course, these considerations leave open a number of fundamental issues in considering such ultra-dense systems. Foremost, how could such systems ever have formed from an ISM? Or, how does stellar evolution proceed in an environment of frequent stellar envelope piercing?

\subsection{Variations in the initial mass function}\label{sec:imf}

\begin{figure*}
     \centering
    \includegraphics[width=0.7\linewidth]{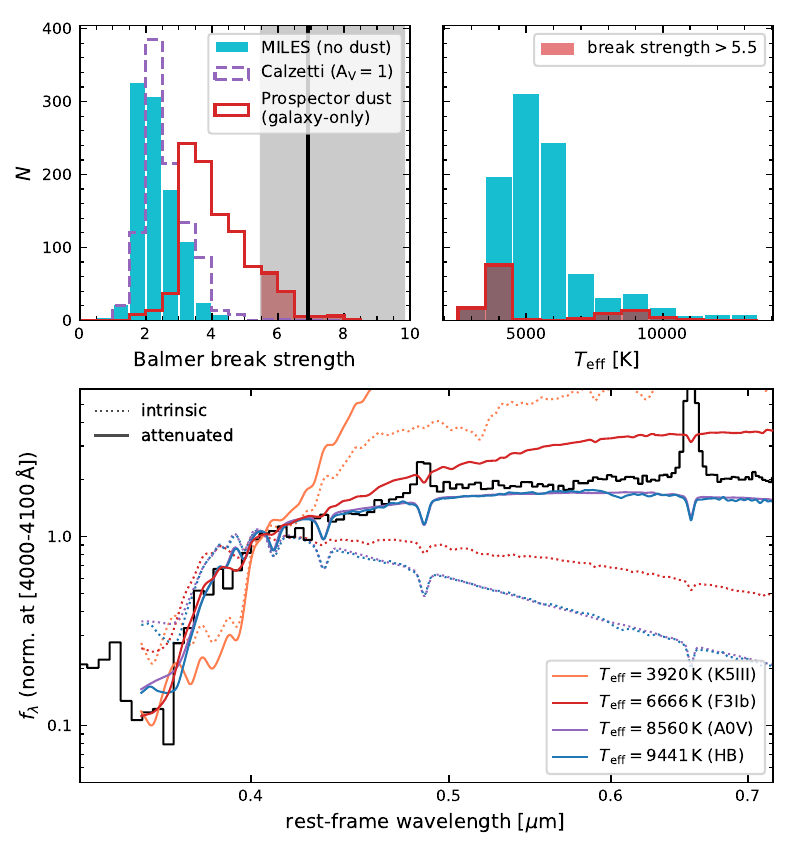}
    \caption{Top left: Balmer break strength (from rest-frame $[3620,3720]\,\AA$ and $[4000,4100]\,\AA$ ranges) distribution for the standard MILES stellar library (blue), and after applying a Calzetti dust law (with $\Av=1$; dashed purple), and the steep attenuation law from the \prospector galaxy-only model (solid red). The black vertical line and shaded region show the measured break strength and $1\sigma$ uncertainty of \ruby. Top right: Effective temperatures of all MILES stars (blue) and the stars that, after applying the steep dust law, fall within $1\sigma$ of \ruby (red). Bottom: A representative sample of MILES spectra with strong Balmer breaks, convolved to the PRISM resolution at $z=3.5$ and normalised at $[4000,4100]\,\AA$. Dotted and solid lines show the stellar spectra without dust and with steep dust attenuation, respectively. In comparison to the spectrum of \ruby, K giants, main-sequence A stars and horizontal branch stars fall short at $<3645\,\AA$ and $>5000\,\AA$. Only massive supergiants, with steep dust attenuation, can match the strength and shape of the Balmer break (albeit with a mismatch at $>5000\,\AA$), and suggests that an extraordinary, top-heavy IMF would be required to explain the rest-frame optical emission of \ruby. }
    \label{fig:miles}   
\end{figure*}

Thus far, we have made the assumption that the IMF of the system resembles that of the Milky Way. At higher redshifts, however, this assumption may no longer hold due to different conditions of the interstellar medium. Top-heavy IMFs have recently been explored as a solution to explain the discovery of extremely massive and extremely UV-luminous galaxies at $z\gtrsim6$ \citep[e.g.][]{Harikane2023:uvlf,Steinhardt2023,Woodrum2024,Cueto2024,Mauerhofer2025}. Such IMFs increase the ratio of massive to dwarf stars and therefore can reduce the mass-to-light ratio ($M/L$). In the context of this work, top-heavy IMFs are interesting because a higher fraction of stars $>1\,\Msun$ could also increase the Balmer break strength.

Exploring variations in the IMF in a self-consistent modelling framework such as \prospector is challenging, and the parameter space to explore (i.e. IMF parametrisation, star formation history, dust law) is potentially extremely large. We therefore take an empirical approach instead, and leverage the fact that, ultimately, any stellar population synthesis model represents a weighted average of a large and diverse set of stellar spectra. 
Moreover, because \ruby shows an extremely strong Balmer break and its blue rest-frame near-infrared SED is suggestive of the Rayleigh-Jeans tail of a black body, it is reasonable to assume that a single component (i.e. a narrow range in stellar effective temperatures) dominates such a weighted average. 

We therefore search for stars that most closely resemble the SED of \ruby in one of the most commonly used stellar libraries for constructing SPS models, MILES \citep{Sanchez-Blazquez2006}, which was also used in our modelling in Section~\ref{sec:sed_models}. To narrow down our search, we compute Balmer break strengths for the spectra using the same definition as in Section~\ref{sec:spec} and further inspect only those stars with strong breaks. The distribution of break strengths is shown in the top left panel of Figure~\ref{fig:miles} (blue). Although the shape of this distribution is arbitrary, i.e., it does not reflect our assumed IMF, this reveals that the maximum possible break strength in the stellar library $\sim 4$, significantly lower than the measured Balmer break of \ruby (black solid line). If we apply a \citet{Calzetti2000} dust law with $\Av=1$ (dashed purple), the break strengths increase slightly, but there are still no stars that match the observed Balmer break. We also apply the steep dust attenuation of the galaxy-only \prospector model from Section~\ref{sec:prospector} (red solid), which increases the typical Balmer break by a factor $\sim2$ and yields a subset of stars with break strengths comparable to \ruby.

Next, we select the strongly-attenuated stars with Balmer breaks $>5.5$, which overlap with the $1\sigma$ uncertainty of \ruby. The distribution of effective temperatures (top right) is bimodal: the majority of stars have $T_{\rm eff}\sim4000\,$K, and a smaller subset is clustered at $T_{\rm eff}\sim8000\,$K. Upon visual inspection, we identify four types of stars and show a representative spectrum for each type in the bottom panel of Figure~\ref{fig:miles} (where we have convolved the MILES spectra to the PRISM resolution at $z=3.5$). The stars with low $T_{\rm eff}$ have the strongest breaks (up to a value of $\sim8$), and are giant stars with strong metal absorption features blueward of $4000\,\AA$, leading to the well-known $4000\,\AA$ break. Although these giant stars do have strong Balmer breaks by the (broad) definition we have used, their spectral shapes clearly do not match the SED of \ruby. 

At $T_{\rm eff}\sim8000-10000\,$K main sequence A stars yield Balmer break strengths $\lesssim 6.5$ after applying steep dust attenuation; horizontal branch stars in this temperature range show marginally stronger breaks. In detail, however, Figure~\ref{fig:miles} shows that these stars are not sufficiently blue at $<3645\,\AA$ and not red enough at $>5000\,\AA$. This is similar to our findings in the SED modelling of Section~\ref{sec:sed_models}. Importantly, it implies that, even if A stars were to dominate the SED entirely, \ruby cannot be modelled as a `normal' post-starburst galaxy.

Intriguingly, we identify one star in the MILES library that matches the Balmer break of \ruby remarkably well (after substantial dust attenuation; red line in Figure~\ref{fig:miles}), although the two spectra start to diverge at $>4500\,\AA$. Its spectral type, F3 Ib, corresponds to a yellow supergiant with $T_{\rm eff}\sim6700\,$K. Such stars represent a brief phase ($\sim 10^{3-4}\,$yr) in the evolution of a massive star (initial mass $\sim 8-20\,\Msun$; e.g. \citealt{Drout2009}). 

This suggests a stellar population with an extraordinarily top-heavy IMF -- coupled with a very steep dust law -- may be able to explain the SED of \ruby. It would, however, require a great degree of fine tuning of the star formation history to achieve the right composition of A and supergiant stars needed to match the spectrum, as the supergiant phase is extremely short-lived and the presence of any O or B stars would wash out the strong Balmer break. Importantly, such an IMF would drastically lower the $M/L$ and hence also reduce the stellar mass density. We estimate the change in $M/L$ (at $5500\,\AA$) for extreme IMFs by constructing simple stellar populations for a range of `$\delta$-function' IMFs in FSPS, attenuated by the same dust curve as the \prospector galaxy-only model. To do so, the IMF is parametrised by two power laws with a steep negative slope ($\gamma_1$) for masses $m<m_0$ and positive slope ($\gamma_2$) for $m>m_0$, with transition mass $m_0=2\,\Msun$ (i.e. main-sequence A stars). For stellar populations dominated entirely by A stars ($\gamma_1=-50$, $\gamma_2=50$, ages $\sim0.1-1\,$Gyr), the $M/L$ is a factor $\sim10$ lower than for the models of Section~\ref{sec:sed_models}. If we allow for a larger fraction of massive stars (by varying $\gamma_2\sim1-2$), we find differences in $M/L$ of a factor $\sim10-50$ depending on the age. The lower mass density of the system in turns implies a considerably lower stellar velocity dispersion, by a factor $\sim 3-7$ ($\sigma_*\sim100-200\,\kms$).

Recent work by \citet{vDokkum2024} proposed that a `ski-slope' IMF (i.e. a steep slope at $<0.5\,\Msun$ and shallow slope at $>1\,\Msun$) could reconcile observations of massive galaxies at high and low redshifts, as such an IMF would look top-heavy at high redshifts due to the higher fraction of massive stars, while a large population of low-mass stars would appear as a bottom-heavy IMF in observations of local early-type galaxies. It may be possible to conceal a large population of low-mass ($<0.5\,\Msun$) stars in \ruby, but the increased number density of both dwarf and giant stars in such a scenario would also substantially increase the cross section of the stellar population. Although a detailed calculation, with revised stellar mass and hence velocity dispersion estimates, is necessary to determine the corresponding change in the stellar collision rate and X-ray luminosity, it is likely that such a system would violate current X-ray constraints.

\begin{figure*}
    \centering
    \includegraphics[width=0.88\linewidth]{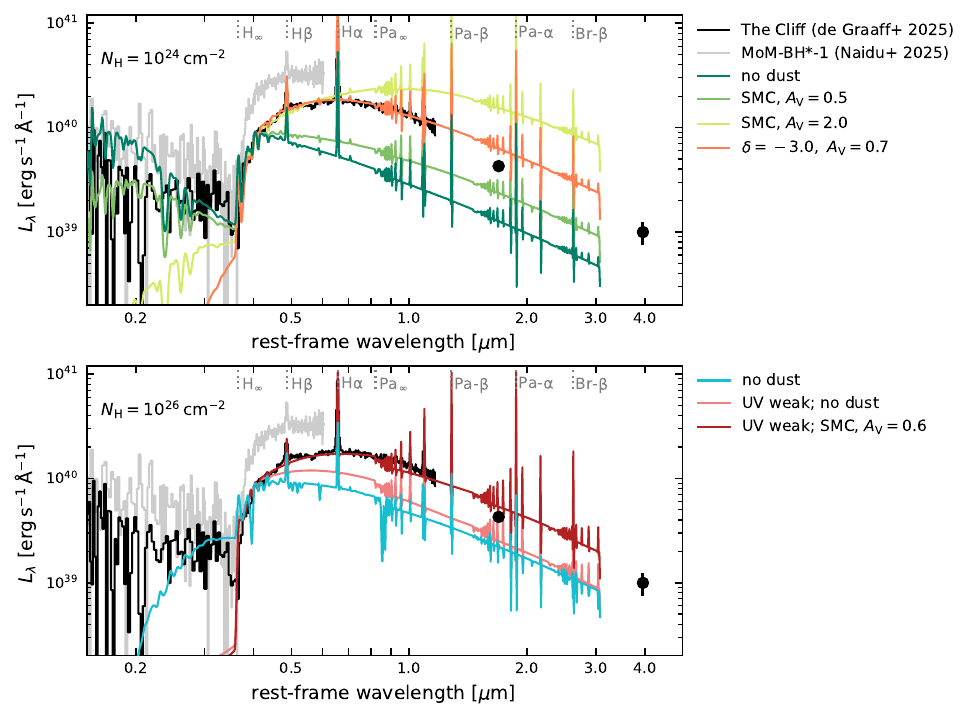}
    \caption{The NIRSpec/PRISM spectra of \ruby and \mombh ($\zspec=7.76$; \citealt{Naidu2025}) are remarkably similar in shape, and only differ by a factor $\sim2$ in luminosity. Top: A BH* star model of a blue incident AGN spectrum ($\alpha_X=-1.5$, $\alpha_{\rm UV}=-0.5$) with high column density of $N_{\rm H}=10^{24}\,$cm$^{-2}$ (rescaled to the rest $[4000,4100]\,\AA$ flux of \ruby and convolved to PRISM resolution) yields an intrinsically strong Balmer break (dark green) that matches the spectrum better than any of the stellar spectra explored (Figure~\ref{fig:miles}). Steep dust curves are required to match the shape of the spectrum blueward of $\sim4200\,\AA$ (green), and high optical depths are needed to be able to match the curvature of the rest-optical spectrum (light green, orange). However, such strongly dust-reddened AGN models result in a severe discrepancy in the rest near-IR. Bottom: Increased absorption from dense gas ($N_{\rm H}=10^{26}\,$cm$^{-2}$) results in extra reddening of the spectra without the need for dust (blue line), but the Balmer break of this model does not match the spectrum. A UV-weak incident AGN spectrum ($\alpha_X=-0.5$, $\alpha_{\rm UV}=-0.1$) can produce a strong Balmer break as well as a redder rest near-IR, although small mismatches in the IR remain even after modest dust attenuation (red lines). This suggests a redder intrinsic AGN spectrum (as predicted by some super-Eddington accretion disc models), or additional component in the SED such as (super)massive stars, is needed to explain \ruby. }
    \label{fig:BHstar}
\end{figure*}

We therefore conclude that it may be theoretically possible to construct a star formation history, dust law, and top-heavy IMF that reproduces the optical to near-IR continuum of \ruby (at a lower stellar mass density), but that the SED resulting from such a contrived setup would be short-lived and therefore extremely rare. Yet, the fact that a second source with near-identical SED shape and luminosity has been found in the same small survey area of $\sim 200\,$arcmin$^2$ (Figure~\ref{fig:BHstar}; \citealt{Naidu2025}) appears to argue against these sources being rare transient phenomena. In the following section, we therefore proceed to discuss alternative models that do not require any (prominent) stellar component in the rest-optical.

\subsection{A `Black hole star'?}\label{sec:cloudy}

In Section~\ref{sec:elines} we showed that the \Ha emission complex of \ruby contains a broad Lorentzian component of $\rm FWHM\sim1500\,\kms$. Although we do not detect any forbidden lines (e.g. \Oiii) and thus lack a strong constraint on the narrow line kinematics, such a broad profile points to the likely presence of an accreting black hole. Assuming the single epoch scaling relation between black hole mass, \Ha luminosity and \Ha line width of \citet{Greene2005}, we obtain a black hole mass of $\log(\Mbh/\Msun)=7.18_{-0.06}^{+0.07}$; as we will discuss further below, we do not include a reddening correction, but note here that $\Mbh$ increases by 0.54\,dex for $A_{\rm H\alpha}=2\,$. The bolometric luminosity implied by the \Ha line luminosity (assuming the scaling relations between $L_{\rm H\alpha}$, $L_{5100}$ and $L_{\rm bol}$ of \citealt{Greene2005} and \citealt{Shen2020}), $L_{\rm bol}\sim1\times10^{45}\,\ergs$, is a factor $\sim 5-10$ higher than the bolometric luminosity implied by the $3\sigma$ upper limit on the soft X-ray luminosity, but consistent with the $3\sigma$ upper limit from hard X-rays (see Section~\ref{sec:xray}).

The high observed equivalent width of the broad \Ha line, $\rm EW_{H\alpha, rest}\sim 400\,\AA$, suggests that AGN emission must also account for a substantial fraction of the rest-optical continuum. However, in Section~\ref{sec:sed_models} we showed that typical AGN models, i.e. blue SEDs described by (piece-wise) power laws, cannot reproduce the spectrum of \ruby even after strong dust attenuation. The difficulty in finding a suitable AGN model is that it needs to fit the Balmer break region, produce a red rest-optical continuum, yet a (comparatively) blue rest-frame near-IR. The absence of X-ray emission as well as hot dust emission in the mid-IR (at least out to rest $4\,\micron$) pose additional challenges.

Interestingly, our exploration of the MILES stellar library in the previous section showed that some stellar spectra (to first order) are a good match to \ruby. \citet{Inayoshi2025} recently proposed that dense gas clouds in the vicinity of an AGN accretion disc could mimic some of the conditions of a stellar atmosphere, such as the high gas density ($n_{\rm H}\sim10^9\,$cm$^{-3}$), resulting in a strong Balmer break and Balmer absorption lines. \citet{Ji2025} implemented such a model in the photoionisation software \cloudy \citep{Ferland2017} and showed that an incident AGN spectrum transmitted through a slab of dense gas with high turbulent velocity can indeed produce both a strong Balmer break and strong Balmer absorption. They hence showed that, after applying an SMC attenuation law with $\Av=2.1$, this results in a reasonable fit to the spectrum of the triply-imaged LRD A2744-QSO1 of \citet[][]{Furtak2024}, with a Balmer break that is slightly \emph{stronger} than observed. Although the Balmer break of A2744-QSO1 (shown in Figure~\ref{fig:spec}, red line) is less strong (by a factor 2) than that of \ruby, the two sources do share some other characteristics, such as luminous and broad Balmer emission, very weak \Oiii emission, and narrow Balmer absorption.

\subsubsection{Dust-reddened BH* models struggle in the infrared}

Such a model, hereafter referred to as a `black hole star' (BH*), may therefore also be a good description of the extremely strong Balmer break and absorption in the \Ha line seen in \ruby. Although this term has some similarities to the `quasi-stars' proposed in the past to explain the rapid growth of light seeds into intermediate mass black holes \citep[e.g.][]{Begelman2006,Begelman2008}, the BH* model does not necessarily imply spherical symmetry, hydrostatic equilibrium, or require the collapse of pristine gas. During the writing of this paper \citet[][hereafter N25]{Naidu2025} discovered a source (named \mombh) with a near-identical spectral shape to \ruby in the same imaging area of the UDS, but at $\zspec=7.76$. We show in Figure~\ref{fig:BHstar} that these sources differ only by a factor $\sim2$ in luminosity, although the Balmer emission of \mombh is a factor $\sim 2$ broader in width. As discussed in detail in N25, the spectrum of \mombh can be well described with an AGN-dominated model in which a shell of dense gas (i.e., with a high covering fraction, for simplicity set to 100\%) surrounds the accretion disc of a massive black hole. In what follows, we will use the modelling framework presented in their work. Briefly, the free parameters of this model consist of the gas density ($n_{\rm H}$), column density ($N_{\rm H}$), metallicity ($\log Z$), ionisation parameter ($\log U$), and the turbulent velocity ($v_{\rm turb}$); the AGN accretion disc spectrum uses the default AGN models of \cloudy, parametrised by the temperature of the big blue bump and three power laws. The transmitted spectrum is then computed with \cloudy, where in practice the 3D spherical shell model is simplified to a slab of gas, and both the continuum and line emission from the dense gas itself are taken into account. Lastly, we note that, because only a finite number of Hydrogen energy levels are computed, there is an artificial emission line at $\sim3650\,\AA$ from the blending of higher order transitions.

The parameter space of these BH* models is high-dimensional, and directly fitting for the best model is at present not feasible. We therefore start by examining a BH* model that is broadly similar to the one proposed by \citet{Ji2025}, with $n_{\rm H}=10^{10}\,$cm$^{-3}$, $N_{\rm H}=10^{24}\,$cm$^{-2}$, $\log(Z/Z_\odot)=-1.5$ and $\log U = -1.5$. A high turbulent velocity ($v_{\rm turb}=500\,\kms$) is needed to produce the EWs of the Balmer absorption features seen in \mombh. Whether this is also needed for \ruby is unclear, primarily due to the poorly constrained absorption feature in the G395M spectrum (Section~\ref{sec:elines}). For the purpose of this paper we therefore focus only on the shape of the continuum and defer a detailed modelling of the emission lines and Balmer absorption to a future study with deeper G395M data (GO-7488; PI: A. Weibel). The incident AGN spectrum is described by $T=10^5\,$K and $\alpha_{\rm X}=-1.5$, $\alpha_{\rm UV}=-0.5$, $\alpha_{\rm ox}=-1.5$ for the X-ray, UV and X-ray to optical power-law slopes, respectively. This model was drawn from a large grid ($\sim10^6$ models) and chosen to produce a qualitatively similar spectral shape (Balmer break strength, UV and optical continuum slope) as \citet{Ji2025}. The model is normalised to the spectrum of \ruby by the luminosity at rest $[4000,4100]\,\AA$ and convolved to the PRISM resolution at $z=3.55$.

The intrinsic BH* model (top panel of Figure~\ref{fig:BHstar}; dark green line) has a Balmer break strength $\sim 4$,  stronger than that of any star with $T_{\rm eff}>6000\,$K explored in Section~\ref{sec:imf}, and therefore performs better than the stellar population models attempted before. However, the model slightly overpredicts the rest-UV emission, and severely underpredicts the emission in the rest-optical and near-IR. We next apply an average SMC dust law with the low optical depth observed in the SMC ($\Av=0.5$; \citealt{Gordon2003}). This model (top panel, green line) provides a remarkably good fit in the rest-frame UV and up to rest $\sim4200\,\AA$, but falls short at longer wavelengths. Following \citet{Ji2025} we also show the SMC attenuation with $\Av=2$, although we caution that this dust law suffers from exactly the same problems discussed in Section~\ref{sec:dust}, as the steepening of the SMC dust law has been shown to arise from a specific star-dust geometry at low optical depth (e.g. see \citealt{Chevallard2013,Salim2020}). The light green line shows this attenuated model, providing an excellent match to the region around the Balmer break ($\sim3645-4800\,\AA$), and a reasonable fit out to \Ha. The rest-UV emission is nearly completely absorbed, although this could be remedied by, e.g., invoking a low-mass star-forming host galaxy outside of the hydrogen envelope, or allowing for scattered light from the AGN by lowering the covering fraction. 

More importantly, this strongly dust-reddened BH* model reveals a critical issue in the rest near-IR, as the flux at wavelengths beyond \Ha deviates significantly from the observed spectrum. This issue is not apparent in either of the NIRSpec spectra modelled by \citet{Ji2025} and N25, as both their objects are at $z\sim7-8$ where NIRSpec provides coverage out to the rest-optical and only MIRI data is able to constrain the rest IR. 
With complete coverage from X-ray wavelengths to the rest mid-IR, \ruby therefore provides a unique test-bed for future BH* models.

\subsubsection{Alternatives to dust reddening}

What is needed in order to fit the SED of \ruby? The fact that the BH* model is able to fit the rest-UV and Balmer break (up to $\sim 4200\,\AA$) with minimal dust attenuation is encouraging, and suggests that this class of models may provide a reasonable description of the underlying physics. However, the discrepancy in the rest-optical and near-IR clearly indicates that some modification is necessary. 
It is possible to achieve extra reddening in the near-IR with a dust law steeper than that of the SMC, as shown in the top panel of Figure~\ref{fig:BHstar} (orange line). But, invoking such a dust law with (comparatively) high optical depth is problematic (see Section~\ref{sec:dust}), and still cannot produce a spectrum that simultaneously matches the rest-optical and near-IR. It may also be difficult to reconcile strong dust attenuation with the non-detection of hot dust in the mid-IR, {and we find that \ruby is also not detected in archival Spitzer/MIPS and Herschel/PACS data (obtained from S. McNulty et al. in prep.)}, although deep far-IR observations would be needed to {robustly} constrain the presence of substantial warm or cold dust. \citet{Setton2025} report simultaneous mid- and far-IR observations for two different luminous LRDs and show that the maximum IR luminosity limits are in tension with models that invoke strong dust attenuation. Similarly, deep far-IR observations of (bright) LRDs at $z\sim3-7$ thus far by \citet{Xiao2025} and \citet{Akins2025} reveal no (cold) dust continuum emission. {Only \citet{Barro2024b} report a compact red source that is well-detected in the mid- and far-IR indicative of hot dust, although this source does not yet have a spectroscopically confirmed v-shaped rest-UV to optical continuum or broad Balmer line, and it is therefore unclear whether this is a LRD.} Under the assumption of energy balance, the BH* model attenuated by an SMC dust law with $\Av=2.0$ would yield an IR luminosity of $10^{11.2}\,L_\odot$, which is only 0.5 dex below the 3$\sigma$ limit on the total IR luminosity for LRDs measured in \citet[][after normalising to rest $6000\,\AA$]{Setton2025}, and thus detectable with ground-based sub-mm facilities.

We therefore propose that the AGN SED may be intrinsically red (at rest-optical wavelengths), rather than dust-reddened. This could be achieved by increased reddening from dense absorbing gas, i.e. by increasing the gas density. The bottom panel of Figure~\ref{fig:BHstar} shows the same BH* model as before, but for a higher density of $n_{\rm H}=10^{11}\,$cm$^{-2}$ and $N_{\rm H}=10^{26}\,$cm$^{-2}$ (blue line). This results in a redder rest-optical and rest near-IR continuum, but the resulting Balmer break is not sufficiently strong. Alternatively (or in addition), the accretion disc spectrum could be redder than that of typical AGN, which has been predicted by some super-Eddington models \citep[e.g.][]{Abramowicz1980,Jiang2014,Jiang2019} and, separately, models of accreting black holes inside dense gaseous envelopes (i.e. quasi-stars) have been linked to super-Eddington accretion \citep[albeit for lower black hole masses, e.g.][]{Volonteri2010,Coughlin2024}. Moreover, super-Eddington accretion has also been suggested as a solution to the observed X-ray weakness of LRDs \citep{Lambrides2024,Yue2024,Ananna2024,Inayoshi2024:xray}, although the presence of extremely dense ($N_{\rm H}=10^{26}\,$cm$^{-2}$), Compton-thick gas in the BH* model may alone be sufficient to explain such X-ray weakness. Using the bolometric luminosity and black hole mass estimated from the broad \Ha line under the assumption that local scaling relations apply to \ruby, we obtain an Eddington ratio $L_{\rm bol}/L_{\rm edd}\approx1.3$. However, as no source like \ruby has been discovered before, it is unclear that commonly-used scaling relations can be applied. The broad wings of the Lorentzian line profile may be due to other processes, such as resonant scattering (see N25) or electron scattering {\citep[][]{Laor2006,Rusakov2025}}, and would imply that the broad line width does not reflect the kinematics of the broad line region. The black hole mass may therefore be severely overestimated (by up to $\sim2\,$dex), which in turn would strongly increase the Eddington ratio ($L_{\rm bol}/L_{\rm edd}\sim10$; N25). 

Two models are shown in the bottom panel of Figure~\ref{fig:BHstar} (red lines) for extremely very high column densities ($N_{\rm H}=10^{26}\,$cm$^{-2}$) and AGN spectra that have shallower X-ray and UV slopes ($\alpha_{\rm X}=-0.5$, $\alpha_{\rm UV}=-0.1$, $\alpha_{\rm ox}=-1.5$), corresponding to the fiducial model parameters of N25. These intrinsically redder models provide a better match to the overall shape of the rest-optical and near-IR continuum. Nevertheless, small yet significant mismatches in the near-IR (rest $\sim0.8-1.2\,\micron$) remain, even after modest dust attenuation (SMC law with $\Av=0.6$). 
In our initial exploration of the large parameter space spanned by the AGN and gas properties, we tested only a limited variety of AGN models (spanning a grid of $10^6$ models). It is therefore possible that a better model does exist in a larger parameter grid, but we defer such a detailed empirical investigation of the intrinsic AGN SED to a future paper. 

The fact that none of the BH* models presented above can match the spectrum of \ruby in detail could also imply that we are missing a key ingredient, such as a stellar component. For instance, AGN models with intrinsically red SEDs have been proposed in the past which do not require super-Eddington accretion rates, but instead impose a heating of the outer accretion disc by (super)massive stars \citep[e.g.][although these models may be too luminous in the mid-IR]{Sirko2003}. More generally, we have so far assumed that an accretion disc powers the BH*, but in principle \emph{any} powerful ionising source inside the dense gaseous envelope could produce a strong Balmer break and luminous emission lines.

Nuclear star clusters, possibly surrounding a massive black hole, may contain such a large population of massive stars with hard ionising spectra: observations of the nuclear star cluster in the Milky Way have revealed an extremely top-heavy IMF \citep{Bartko2010,Lu2013}. If such a nuclear star cluster also contains supermassive stars, which are expected to have modest effective temperatures and hence red spectra \citep[e.g.][]{Martins2020}, this may also help to explain the rest near- and mid-IR SED of \ruby.  Interestingly, the formation of nuclear star clusters and the growth of massive black holes have been demonstrated to be tightly linked, and runaway collisions in nuclear star clusters have long been proposed as a formation channel of massive black holes \citep[for a review, see][]{Neumayer2020}. \citet{Goodman2003} showed that massive stars could even form in the outer accretion disc of a massive black hole, and therefore could be encompassed in the same dense gaseous envelope. To what extent the stellar emission would contribute to the SED, and whether this scenario can produce the broad symmetric Balmer emission lines is currently unclear. Moreover, dedicated models are needed to assess the effects of stellar feedback on the dense gas in such a system and determine whether this would be a short-lived phase or a stable mode of star formation. Considering the high number density of LRDs \citep[e.g.][]{Kokorev2024}, a predominantly AGN origin of the rest-optical to near-IR SED of \ruby therefore appears most plausible at face value, but stellar emission cannot yet be ruled out, and may even contribute substantially.

\section{Conclusions}\label{sec:conclusions}

We present a detailed investigation of \ruby, a luminous LRD at $\zspec=3.55$ with an extremely strong Balmer break discovered in RUBIES. With high-quality JWST/NIRSpec PRISM and G395M spectroscopy from RUBIES, JWST/NIRCam and MIRI imaging from the PRIMER survey, and archival Chandra X-ray data, it is one of few (bright) LRDs with complete spectro-photometric coverage from X-ray to mid-IR wavelengths. Coupled with its highly unusual SED, \ruby provides a crucial stress test of stellar population and AGN models. 

The spectral and morphological properties of \ruby present a complex puzzle:
\begin{itemize}
    \item The SED is characterised by a Balmer break that exceeds the break strengths of $z>3$ massive quiescent galaxies and LRDs published thus far by a factor 2. Together with a (weak) blue UV slope, this gives rise to the `v-shape' typically observed in LRDs. The PRISM spectrum (extending to rest $1.2\,\micron$) and MIRI photometry (rest $\sim 2-4\,\micron$) reveal a blue SED in the near- and mid-IR, indicating there is no evidence of hot dust emission at least to rest $\lesssim4\,\micron$.
    \item The PRISM spectrum reveals a suite of Balmer, Paschen and \Hei emission lines, but no significant metal lines, pointing to low-metallicity or high-density gas. The G395M spectrum kinematically resolves \Ha and various Pa lines, showing broad Lorentzian profiles ($\rm FWHM\sim1500\,\kms$) as well as a narrow redshifted absorption feature in the \Ha line. If we assume this originates from the broad-line region of an AGN and use low-redshift single-epoch black hole mass scaling relations, the \Ha line translates to a black hole mass $\Mbh\sim1\times 10^7\,\Msun$.
    \item S\'ersic profile fitting to NIRCam F200W imaging shows that the source is extremely compact, but possibly marginally resolved compared to a nearby star. The single-component S\'ersic model converges to $\re\sim40\,$pc, with a robust $2\sigma$ upper limit of $\re<51\,$pc.    
    \item We do not detect any significant X-ray emission in either soft or hard X-ray bands (rest $\sim1-10\,$keV and $10-30\,$keV). 
\end{itemize}

We perform spectrophotometric fitting to rigorously test current state-of-the-art stellar population, AGN, and composite models. 
\begin{itemize}
    \item We demonstrate that SED models in which a galaxy (i.e. evolved stellar population) dominates the rest-optical continuum systematically overpredicts the flux blueward, and underpredicts the flux redward of the Balmer break. This remains the case even when including an AGN component in the model, and when allowing for extremely steep dust laws with unusually high optical depths. 
    \item The ill-fitting star-dominated SED models favour a massive ($\log(M_*/\Msun)\sim10^{10.4-10.6}$), reddened post-starburst solution. We show that the implied extremely high stellar mass density ($\rho\sim10^5\,\Msun\,{\rm pc}^{-3}$) would lead to frequent stellar collisions between giant and main-sequence stars with a collision rate of $\sim5\,$yr$^{-1}$, which may produce detectable bursts of X-ray emission.
    \item By exploring stellar spectra, we demonstrate that main-sequence A stars alone cannot produce the Balmer break and shape of the rest-optical SED, even after attenuation with steep dust laws. Altering the IMF slope to boost the relative number of A stars is thus not sufficient to explain the SED. Only an extremely top-heavy IMF with a significant fraction of supergiants  -- in combination with a strong and steep dust attenuation curve, and a highly contrived star formation history -- can simultaneously produce a strong Balmer break and red rest-optical continuum. But, we estimate that such an IMF would also lower the stellar mass density by $>1\,$dex, and reduce the stellar velocity dispersion by a factor $>3$. 
\end{itemize}

We therefore conclude that the rest-optical and near-IR continuum of \ruby cannot originate from a massive, evolved stellar population with an extremely high stellar density. Crucially, this is the first LRD for which such a strong conclusion can be drawn solely by analysing the SED.

Instead, we argue that the most plausible model is that of a luminous ionising source reddened by dense, absorbing gas in its close vicinity. 
Currently the only model capable of producing both the strength and shape of the observed Balmer break is that of a `black hole star', in which an AGN accretion disc is embedded in turbulent dense gas. Such a model would also naturally explain the observed absorption in the \Ha line. The BH* models are still in an early phase of development. 
We show that recently proposed dust-reddened BH* models (with $n_{\rm H}=10^{10}\,$cm$^{-3}$ and $N_{\rm H}=10^{24}\,$cm$^{-2}$) can provide a good fit to the rest-optical, but severely overpredict the rest near- and mid-IR.
We therefore propose that the AGN spectrum may be intrinsically red, i.e. intrinsically luminous in the rest-optical in addition to strong reddening from very dense gas ($n_{\rm H}\sim10^{11}\,$cm$^{-3}$; $N_{\rm H}\sim10^{26}\,$cm$^{-2}$), and show that such a BH* model currently provides the best match to the rest-optical to mid-IR of \ruby. Possibly, an intrinsically red spectrum could be explained by less conventional accretion disc models, for instance by invoking super-Eddington accretion (which in turn would imply that our black hole mass estimate is highly uncertain). Alternatively, (super)massive stars formed within the dense gaseous envelope may also contribute to the red continuum emission, but models for such a nuclear star cluster scenario are yet to be explored.

\ruby presents the strongest direct evidence to date that the Balmer break and rest optical to near-IR SED in LRDs can be dominated by emission from an AGN, rather than evolved stellar populations, although many open questions regarding the black hole and host galaxy properties remain. Because of its comparatively modest redshift, the high-quality spectrophotometric coverage of JWST extends over a wide rest-frame wavelength range. These stringent constraints make \ruby the ideal benchmark for future AGN and BH* models.

\begin{acknowledgements}

We thank the PRIMER team for making their imaging data publicly available immediately. We thank Jaime Villase\~{n}or and Friedrich R\"opke for helpful discussions. 
This work is based on observations made with the NASA/ESA/CSA James Webb Space Telescope. The data were obtained from the Mikulski Archive for Space Telescopes at the Space Telescope Science Institute, which is operated by the Association of Universities for Research in Astronomy, Inc., under NASA contract NAS 5-03127 for JWST. These observations are associated with programs \#1837 and \#4233.
Support for program \#4233 was provided by NASA through a grant from the Space Telescope Science Institute, which is operated by the Association of Universities for Research in Astronomy, Inc., under NASA contract NAS 5-03127. REH acknowledges support by the German Aerospace Center (DLR) and the Federal Ministry for Economic Affairs and Energy (BMWi) through program 50OR2403 `RUBIES'.
This research was supported by the International Space Science Institute (ISSI) in Bern, through ISSI International Team project \#562.
The Cosmic Dawn Center is funded by the Danish National Research Foundation (DNRF) under grant \#140. This work has received funding from the Swiss State Secretariat for Education, Research and Innovation (SERI) under contract number MB22.00072, as well as from the Swiss National Science Foundation (SNSF) through project grant 200020\_207349. Support for this work for RPN was provided by NASA through the NASA Hubble Fellowship grant HST-HF2-51515.001-A awarded by the Space Telescope Science Institute, which is operated by the Association of Universities for Research in Astronomy, Incorporated, under NASA contract NAS5-26555. TBM was supported by a CIERA fellowship.

 \end{acknowledgements}

\bibliographystyle{aa}
\bibliography{rubies}

\appendix

\section{JWST/NIRCam and MIRI photometry}\label{sec:apdx_photometry}

\begin{figure*}
    \centering
    \includegraphics[width=\linewidth]{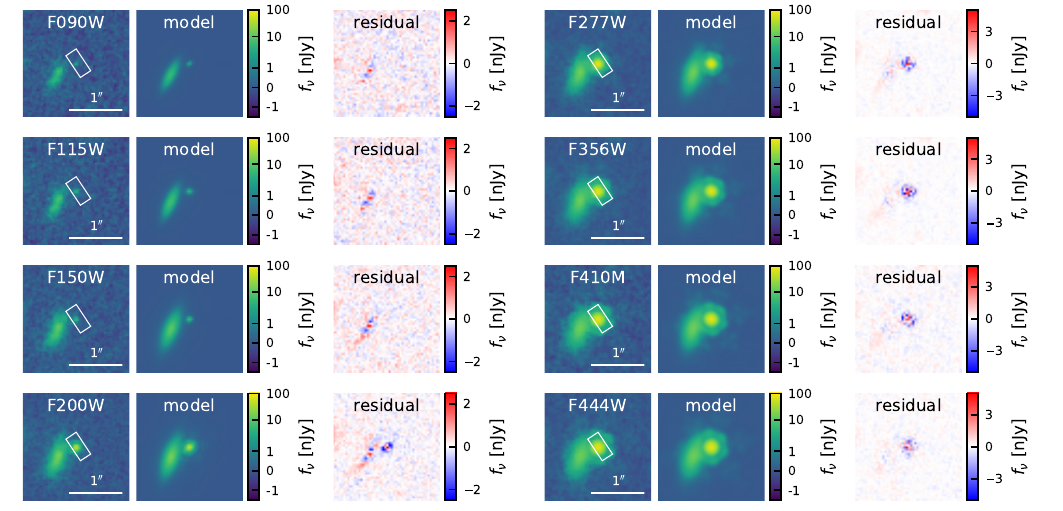}
    \caption{{JWST/NIRCam images ($0.04\arcsec\,{\rm pix}^{-1}$) of \ruby and its nearby lower-redshift neighbour for all available filters. We show the MAP models from the S\'ersic profile fitting for each filter as well as the residuals. The central NIRSpec shutter is plotted for reference.}}
    \label{fig:pysersic_nircam}
\end{figure*}

\begin{figure}
    \centering
    \includegraphics[width=\linewidth]{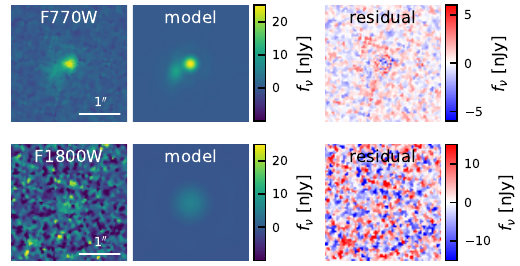}
    \caption{{JWST/MIRI images of \ruby and its nearby lower-redshift neighbour for all available filters. We show the MAP models from the point source $+$ S\'ersic profile fitting for each filter as well as the residuals. }}
    \label{fig:pysersic_miri}
\end{figure}

\begin{figure}
    \centering
    \includegraphics[width=\linewidth]{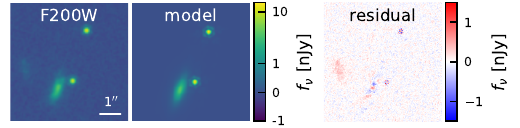}
    \caption{{JWST/NIRCam F200W image at $0.02\arcsec\,{\rm pix}^{-1}$ of \ruby, its nearby lower-redshift neighbour, and the nearby star. We show the MAP model as well as the residuals. }}
    \label{fig:pysersic_20mas}
\end{figure}

\begin{figure}
    \centering
    \includegraphics[width=\linewidth]{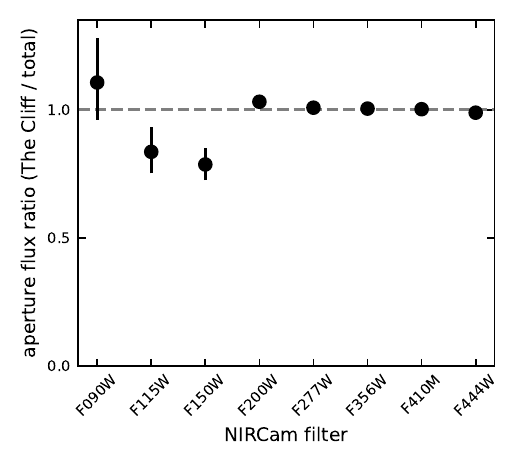}
    \caption{{Ratio between the flux measured from the S\'ersic model of \ruby and the observed NIRCam image, within the rectangular aperture of the NIRSpec microshutter. Error bars reflect both the uncertainty in the S\'ersic fitting and the exact pointing location of JWST. Emission from \ruby dominates ($>80\%$) the total observed flux at all wavelengths, and accounts for $\sim 100\%$ of the total emission redward of the Balmer break ($>2\micron$).} }
    \label{fig:aperflux}
\end{figure}

{The NIRCam images (at pixel scale of $0.04\arcsec\,{\rm pix}^{-1}$), MAP models from \texttt{pysersic} (Section~\ref{sec:imaging}), and image residuals (computed as data$-$model) are shown in Figure~\ref{fig:pysersic_nircam} for all 8 NIRCam filters. Figure~\ref{fig:pysersic_20mas} additionally presents the F200W image, model and residual for a higher pixel scale of $0.02\arcsec\,{\rm pix}^{-1}$ used in Section~\ref{sec:morph}. Finally, Figure~\ref{fig:pysersic_miri} shows the MIRI image cutouts and results from the morphological fitting. We find that the S\'ersic profile fitting performs well across all wavelengths, with only minor residual features remaining.

We use these models to determine the total integrated flux of \ruby, presented in Table~\ref{tab:photometry}. Moreover, we quantify to what extent the flux at the position of the NIRSpec microshutter aperture originates from \ruby versus the foreground source. To do so, we first measure the shutter aperture flux from the individual NIRCam images. We account for uncertainty in the pointing accuracy of JWST by applying random spatial shifts, using 500 draws from a Gaussian distribution with a dispersion of $0.025\arcsec$ \citep{RigbyJWST}. Next, we perform the same aperture measurements (including the same spatial shifts) for the model image of \ruby alone; to at the same time account for uncertainties in the morphological fitting, we generate 500 model images drawn from the posteriors of the \texttt{pysersic} fits. The ratio of these aperture fluxes is shown for the 8 NIRCam filters in Figure~\ref{fig:aperflux}. \ruby clearly dominates the emission across all wavelengths ($>80\%$), and accounts for 100\% of the flux redward of the Balmer limit. 

We therefore conclude that there is no contamination from the foreground source within the NIRSpec aperture at $>2\,\micron$. At $<2\,\micron$ there may be slight contamination (at the $<20\%$ level), but this does not significantly affect any of the modelling or conclusions -- which rely on the flux at $>2\,\micron$ -- in this work. The Balmer break strength could be even higher than measured from the spectrum, by at most 25\%, but this is subdominant to the large measurement uncertainty shown in Figure~\ref{fig:BB_zspec}.
}

\begin{table}[h]
\caption{NIRCam and MIRI photometry of \ruby obtained with \texttt{pysersic} (Section~\ref{sec:imaging}).}
%\begin{center}
\centering
 \begin{tabular}{ll}\hline\hline
  filter & $f_\nu$ ($\mu$Jy) \\ \hline
  NIRCam/F090W &  $0.0148_{-0.0019}^{+0.0024}$ \\[2pt]
  NIRCam/F115W &  $0.028_{-0.004}^{+0.004}$ \\[2pt]
  NIRCam/F150W &  $0.0269_{-0.0017}^{+0.0022}$ \\[2pt]
  NIRCam/F200W &  $0.322_{-0.028}^{+0.029}$ \\[2pt]
  NIRCam/F277W &  $1.090_{-0.004}^{+0.004}$ \\[2pt]
  NIRCam/F356W &  $1.334_{-0.006}^{+0.005}$ \\[2pt]
  NIRCam/F410M &  $1.597_{-0.013}^{+0.011}$ \\[2pt]
  NIRCam/F444W &  $1.725_{-0.013}^{+0.012}$ \\[2pt]
  MIRI/F770W &  $1.54_{-0.03}^{+0.03}$ \\[2pt]
  MIRI/F1800W &  $1.96_{-0.5}^{+0.5}$ \\[2pt]  \hline
 \end{tabular}
\label{tab:photometry}
%\end{center}
\end{table}

\end{document}